\documentclass[aps,prl,reprint,amsmath,amssymb,superscriptaddress,floatfix]{revtex4-1}
\usepackage{graphicx}
\usepackage[T1]{fontenc}
\usepackage[utf8]{inputenc}
\usepackage{hyperref}
\usepackage[version=4]{mhchem}
\usepackage{gensymb}
\usepackage{siunitx}
\usepackage{letltxmacro}
\usepackage{xcolor}
\usepackage{booktabs}
\usepackage{tabularx}
\usepackage{lineno}

\newcommand{\tred}[1]{\textcolor{black}{#1}}
\newcommand{\tredgram}[1]{\textcolor{black}{#1}}

\renewcommand{\vec}[1]{\mathbf{#1}}
\renewcommand{\k}{\vec{k}}
\newcommand{\q}{\vec{q}}

\renewcommand{\S}{\vec{S}}
\LetLtxMacro{\originaleqref}{\eqref}
\renewcommand{\eqref}{Eq.~\originaleqref}
\newcommand{\figref}[1]{Fig. \ref{#1}}
\newcommand{\figsref}[1]{Figs. \ref{#1}}

\newcommand{\ADDCQM}{\affiliation{Center for Quantum Materials, Seoul National University, Seoul 08826, Republic of Korea}}
\newcommand{\ADDSNU}{\affiliation{Department of Physics and Astronomy, Seoul National University, Seoul 08826, Republic of Korea}}
\newcommand{\ADDIBS}{\affiliation{Center for Correlated Electron Systems, Institute for Basic Science, Seoul 08826, Republic of Korea}}
\newcommand{\ADDOU}{\affiliation{Research Institute for Interdisciplinary Science, Okayama University, Okayama, Okayama 700-8530, Japan}}
\newcommand{\ADDOPU}{\affiliation{Department of Physics and Electronics, Osaka Prefecture University, Sakai, Osaka 599-8531, Japan}}
\newcommand{\ADDISIS}{\affiliation{ISIS Pulsed Neutron and Muon Source, STFC Rutherford Appleton Laboratory, Didcot, Oxfordshire, OX11 0QX, United Kingdom}}
\newcommand{\ADDPSI}{\affiliation{Laboratory for Neutron Scattering and Imaging, Paul Scherrer Institut, CH-5232 Villigen, Switzerland}}
\newcommand{\ADDPCSIBS}{\affiliation{Center for Theoretical Physics of Complex Systems, Institute for Basic Science (IBS), Daejeon 34126, Korea}}

\begin{document}

\title{Momentum-dependent magnon lifetime in metallic non-collinear triangular antiferromagnet \ce{CrB2}}

\author{Pyeongjae Park}
\ADDCQM
\ADDIBS
\ADDSNU

\author{Kisoo Park}
\ADDIBS
\ADDSNU

\author{Taehun Kim}
\ADDCQM
\ADDIBS
\ADDSNU

\author{Yusuke Kousaka}
\ADDOPU

\author{Ki Hoon Lee}
\ADDIBS
\ADDSNU
\ADDPCSIBS

\author{T. G. Perring}
\ADDISIS

\author{Jaehong Jeong}
\ADDCQM
\ADDIBS
\ADDSNU

\author{Uwe Stuhr}
\ADDPSI

\author{Jun Akimitsu}
\ADDOU

\author{Michel Kenzelmann}
\ADDPSI

\author{Je-Geun Park}\email{jgpark10@snu.ac.kr}
\ADDCQM
\ADDIBS
\ADDSNU

\begin{abstract}
Non-collinear magnetic order arises for various reasons in several magnetic systems and exhibits interesting spin dynamics. Despite its ubiquitous presence, little is known of how magnons, otherwise stable quasiparticles, decay in these systems, particularly in metallic magnets. Using inelastic neutron scattering, we examine the magnetic excitation spectra in a metallic non-collinear antiferromagnet \ce{CrB2}, in which Cr atoms form a triangular lattice and display incommensurate magnetic order. Our data show intrinsic magnon damping and continuum-like excitations that cannot be explained by linear spin wave theory. The intrinsic magnon linewidth $\Gamma(\q,E_\q)$ shows very unusual momentum dependence, which our analysis shows to originate from the combination of two-magnon decay and the Stoner continuum. By comparing the theoretical predictions with the experiments, we identify where in the momentum and energy space one of the two factors becomes more dominant. Our work constitutes a rare comprehensive study of the spin dynamics in metallic non-collinear antiferromagnets. It reveals, for the first time, definite experimental evidence of the higher-order effects in metallic antiferromagnets.
\end{abstract}

\maketitle
Understanding spin dynamics is crucial for the studies of magnetism in condensed matters as it determines \tredgram{their} magnetic properties at a microscopic level. Spin dynamics and corresponding magnetic properties in most condensed matters have been commonly described by linear spin-wave theory (LSWT), which interprets collective magnetic excitations as a set of non-interacting quantized quasiparticles, magnons. However, this well-established picture has been known to break down in some cases. For example, non-collinear magnetic systems host exotic features beyond LSWT in their spin dynamics such as higher-order interactions between magnons \cite{Chernychev_review} or magnon-phonon hybridization~\cite{YMnO3_nat_comm}. As far as these phenomena are concerned, triangular lattice antiferromagnets (TLAFs) have been at the center of active research, since they naturally host non-collinear magnetic structure due to their inherent geometrical frustration~\cite{TLAF_review}. Several previous studies have demonstrated such effects in insulating TLAFs with non-collinear magnetic ground states, e.g., magnon linewidth broadening and magneto-elastic modes in hexagonal manganites \ce{\it{R}MnO3} ($R$ = Y, Lu, Ho)~\cite{LuMno3_PRL_EPG, YMnO3_nat_comm, MP_coupling_Petit, HoMnO3_EPG} and delafossites \ce{\it{A}CrO2} ($A$ = Li, Cu)~\cite{CuCrO2_PRB, LiCrO2_Toth}, and renormalization of magnons in $S=1/2$ TLAF \ce{Ba3CoSb2O9}~\cite{BCSO_J.Ma, BCSO_nat_phys}, to name only a few.

While these works have highlighted some of the new features in insulating systems, spin dynamics remains still unexplored in metallic non-collinear magnets. Though the microscopic nature of magnetism in metallic systems can be quite different from that of insulators, the spin dynamics in most metallic magnets has been nonetheless explained by the isotropic Heisenberg model as an approximate description \cite{Itinerant_Heisenberg}. Thus, it comes down to an exciting question of whether such a higher-order interaction can still be present in the spin Hamiltonian even for metallic magnets or not. Moreover, metallic non-collinear antiferromagnets have recently attracted considerable interest due to their unique topological properties, as in \ce{Mn3\it{X}} ($X$=Ge, Sn)~\cite{Mn3Sn_AHE, Mn3Ge_AHE, AHE_Kubler, Mn3Sn_npj}. Therefore, it is timely and urgent to understand the spin dynamics of metallic non-collinear antiferromagnets in depth. Unfortunately, however, there have been very few comprehensive studies on the spin dynamics of metallic non-collinear antiferromagnets using inelastic neutron scattering (INS). It is mainly due to the lack of suitable metallic non-collinear magnetic materials that can be grown into large single crystals for INS.

\begin{figure}[t]
\centering
\includegraphics[width=\columnwidth]{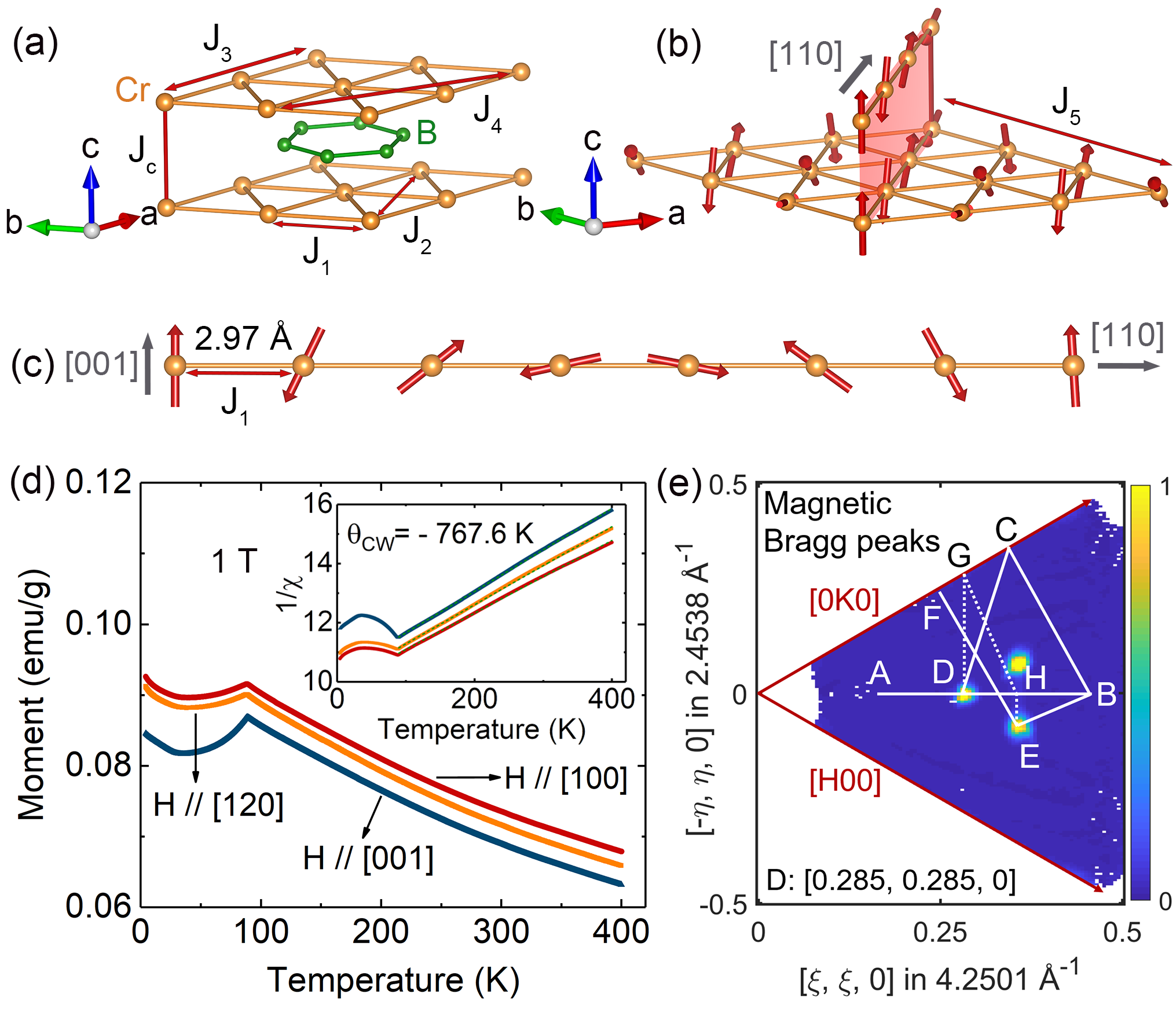}
\caption{(a) Crystal and (b) non-collinear magnetic structure of \ce{CrB2} with up to the $5^{\textrm{th}}$ nearest-neighbor coupling $J_m$ and an interlayer coupling $J_c$ being marked. (c) Incommensurate cycloidal magnetic order of \ce{CrB2} along the [110] direction. (d) Magnetization of single crystal \ce{CrB2} along several directions measured under the magnetic field of \SI{1}{\tesla}. Inset shows inverse susceptibility consistent with the Curie-Weiss behavior above $T_{\textrm{N}}$. (e) Three symmetrically-equivalent magnetic Bragg peaks of \ce{CrB2} with labels of $\q$-points used in \figref{fig:INS}. White solid (dashed) lines indicate the momentum contours used in Fig. 2a (2b), respectively.}
\label{fig:basics}
\end{figure}

Chromium diboride, \ce{CrB2}, is an excellent candidate for this purpose. It has a layered structure ($P$6/$mmm$) with alternating layers of a Cr triangular lattice and a B honeycomb lattice~(\figref{fig:basics}a). It shows long-range non-collinear magnetic order below $T_{\textrm{N}} = \SI{88}{\K}$. The ideal Curie-Weiss behavior with 
$\theta_{\textrm{CW}} \sim -\SI{750}{\K}$ implies sizable frustration among the Cr spins (\figref{fig:basics}d). Earlier neutron diffraction experiments on \ce{CrB2} single crystal revealed cycloidal magnetic order with a propagation vector of $\q_{\textrm{m}}$ = (0.285, 0.285, 0) (\figref{fig:basics}b, c), together with the other two symmetrically equivalent magnetic domains~\cite{NDold_CrB2, NDnew_CrB2} (\figref{fig:basics}e). Although the origin of the incommensurate magnetic order was not fully explained, it is likely to originate from the frustration effect in the exchange coupling due to the presence of further nearest-neighbor interactions~\cite{TLAF_phase_diagram}. Also, \ce{CrB2} has a reduced ordered magnetic moment (0.5 $\mu_{\text{B}}$/f.u.), which is a characteristic feature of itinerant magnets. Finally, temperature dependence of resistivity~\cite{Bauer_CrB2} clearly demonstrates the metallicity of \ce{CrB2}.

In this letter, we report the magnon spectra of a metallic non-collinear antiferromagnet \ce{CrB2} over a wide momentum ($\q$)-energy ($E$) region. By taking advantage of the vast data set, we determine the effective spin Hamiltonian that describes most of the measured spectra. From this analysis, however, we also find unique features that cannot be explained by LSWT: intrinsic magnon damping and additional continuum-like excitations above the magnon branch. To gain further insights, we analyze the intrinsic magnon linewidth at several $\q$ positions, which shows unusual $\q$-dependence. Finally, we discuss the origin of the observed magnon decay and the continuum-like signal by comparing the data with the two-magnon and the Stoner continuum density of states (DOS).

Single crystals of $^{11}$B-enriched \ce{CrB2} were synthesized using the laser floating zone technique to avoid strong neutron absorption of $^{10}$B~\cite{SM}. \figref{fig:basics}d shows the magnetization curves of our sample which are consistent with previous studies~\cite{Bauer_CrB2, MT_CrB2}. Using five co-aligned single crystals with a total mass of \SI{13}{\gram} and the overall mosaicity less than \SI{1.5}{\degree}, we carried out an INS experiment in the MAPS time-of-flight spectrometer at ISIS, UK~\cite{MAPS_ref, MAPS_data_number}. The INS data were collected at $T=\SI{5}{\K}$ with several incident neutron energies of $E_i=40$, $70$, and \SI{150}{\meV}. The data were subsequently symmetrized into the irreducible Brillouin zone 
and analyzed using the Horace software~\cite{Horace_ref}. Background signal and phonon spectra of \ce{CrB2} were subtracted by measuring the empty sample holder separately and with the aid of the DFT calculation, respectively~\cite{SM}. Additional INS data were collected using the EIGER triple-axis spectrometer at PSI, Switzerland with $T= \SI{1.5}{\K}$ and the scattered neutron energy of $E_f= \SI{34}{\meV}$.

\begin{figure*}[t]
\centering
\includegraphics[width=0.8\textwidth]{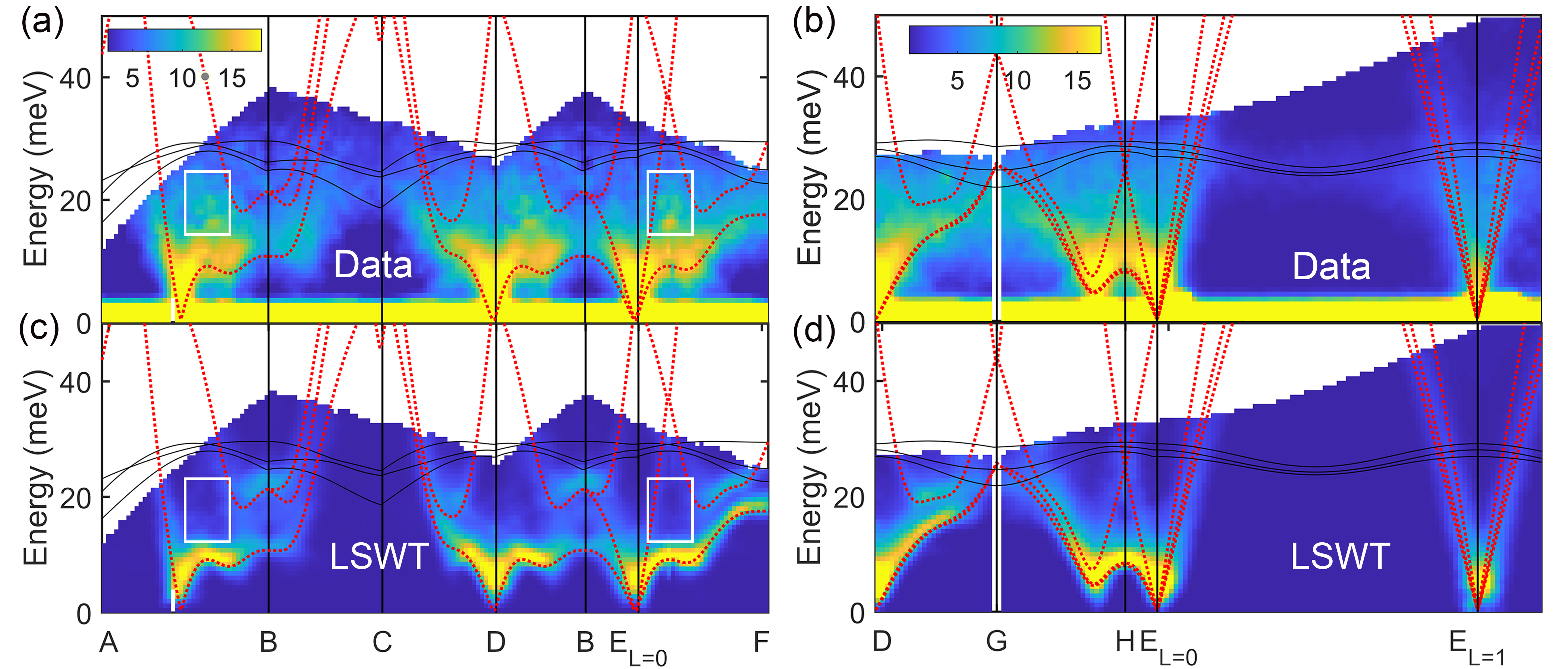}
\caption{(a)-(b) INS data obtained at MAPS along the two paths shown in Fig. 1e, and (c)-(d) simulations of the same spectra using LSWT convoluted with the instrumental resolution of MAPS. Black solid lines are the phonon dispersion lines from the DFT calculations, and red dotted lines are the fitted spin-wave dispersion lines from our magnetic Hamiltonian. White squares highlight the region in the $\q$-$E$ space where there exist continuum-like excitations, which cannot be explained by LSWT.}
\label{fig:INS}
\end{figure*}

\figsref{fig:INS}a and \ref{fig:INS}b show the energy-momentum slices from our INS data along the momentum contours indicated in \figref{fig:basics}e. No magnon energy gap is seen in the INS data, implying negligible anisotropy in \ce{CrB2} consistent with the nearly isotropic magnetic susceptibility. Also, the very steep magnon modes along the [00L] direction ($\textrm{E}_{\textrm{L=0}}$ to $\textrm{E}_{\textrm{L=1}}$ in \figref{fig:INS}b) indicate the three-dimensional nature of magnetism in \ce{CrB2} 
with a sizable interlayer coupling $J_c$.
To explain the observed spin-wave spectra, we adopted the following spin Hamiltonian:
\begin{equation}
\mathcal{H} = \sum_{i,m,j_m} {J_{m}\ \S_{i} \cdot \S_{j_{m}}} + \sum_{i,j_c} {J_{c}\ \S_{i} \cdot \S_{j_{c}}},
\label{eq:Ham}
\end{equation}
%
where $J_m$ and $\S_{j_m}$ denote the coupling constant and the spin moment for the $m^{\textrm{th}}$ in-plane nearest-neighbors, respectively.
%
%
Using the SpinW library~\cite{SpinW_ref}, we calculated magnon dispersion and an INS cross-section from Eq.~\ref{eq:Ham}, including the effect of three symmetrically equivalent magnetic domains. The best fit of the coupling constants $J_m$ and $J_c$ is shown in Table~\ref{table:1}. As one can see, we need the exchange interaction up to the fifth nearest-neighbor, which is due to the long-ranged Ruderman-Kittel-Kasuya-Yosida (RKKY) interaction in metallic magnets (see Fig. S3~\cite{SM}).
\begin{table}[!h]
 \caption{Fitted exchange parameters of \ce{CrB2} with their uncertainty and their coupling distances.}
 \begin{center}
 \centering
 \begin{tabularx}{\columnwidth}{c c c c c c c} 
 \midrule \midrule
  & $J_1S$ & $J_2S$ & $J_3S$ & $J_4S$ & $J_5S$ & $J_cS$ \\
 \midrule
 \ \ $JS$ (\si{\meV}) \ \ & 40.52 & 7.856 & 1.965 & -1.807 & -1.216 & -9.043 \\
 \ \ $\Delta (JS)$ (\si{\meV}) \ \ & ($\pm$)1.62 & 0.608 & 0.079 & 0.061 & 0.122 & 0.452 \\ 
 \ \ Distance (\si{\angstrom}) \ \ & 2.97 & 5.144 & 5.94 & 7.858 & 8.91 & 3.07 \\
 \midrule \midrule
 \end{tabularx}
 \end{center}
\label{table:1}
\end{table}

\begin{figure}[!t]
\centering
\includegraphics[width=\columnwidth]{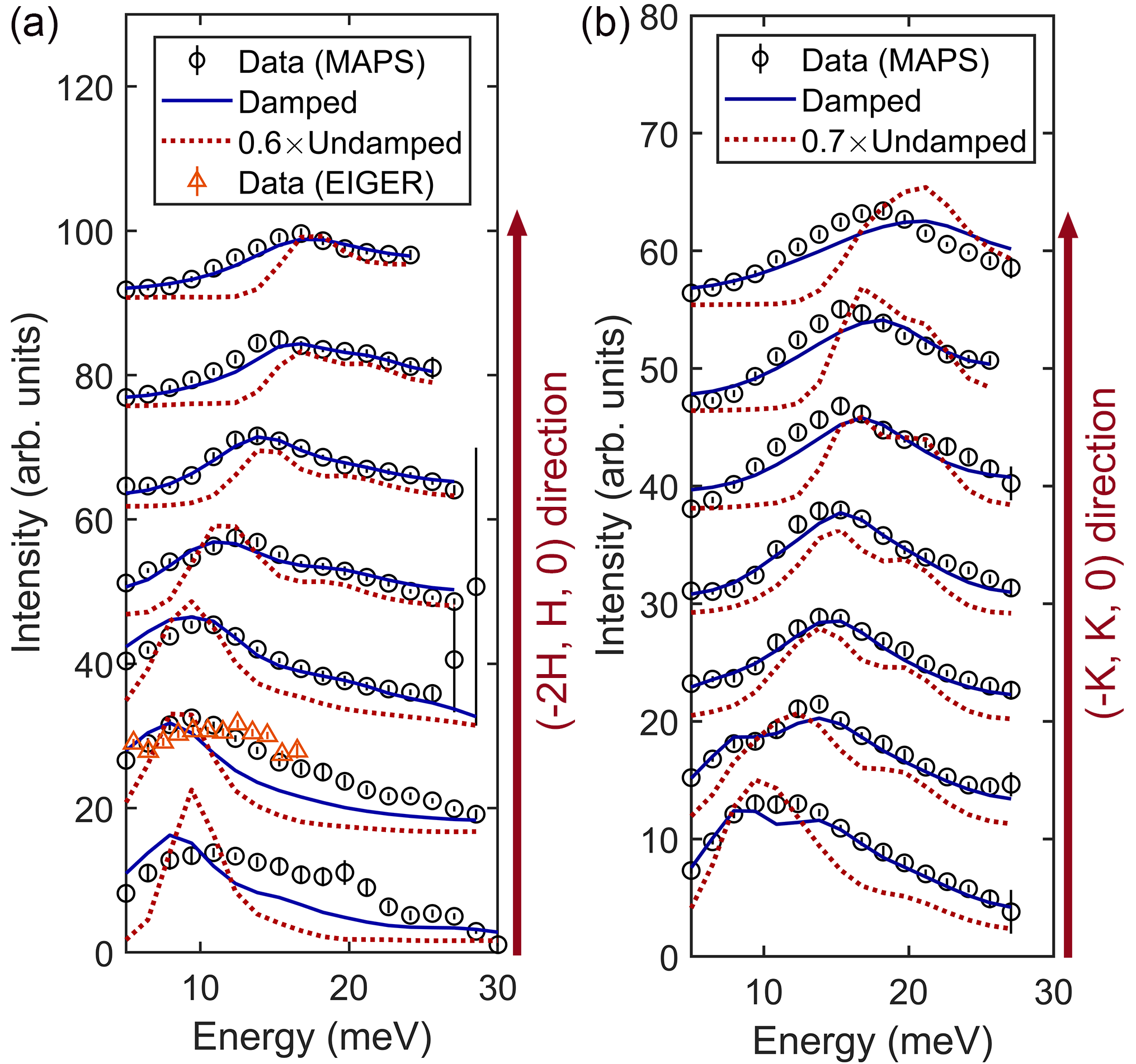}
\caption{(a,b) Comparison between INS data and LSWT simulation with (blue solid lines) and without (red dotted lines) the magnon damping effect considered. (a) and (b) shows a stack of constant $\q$-cuts at different $\q$ position along the [-2H H 0] and [-K K 0] directions, respectively. Each one corresponds to the $\textrm{E}_{\textrm{L=0}}$-F and D-G lines in \figref{fig:INS}. For a better presentation, scale factors were applied to the undamped calculation. The two constant $\q$-cuts at the bottom of (a) clearly show additional continuum-like excitations at the $\q$ positions denoted by the white squares in \figref{fig:INS}a. Error bars indicate the standard deviations of the data points.}
\label{fig:Gamma}
\end{figure}

Although overall features of the magnon dispersion can be described at the LSWT level, there are also distinct signatures that cannot be explained by the theory, calling for something beyond LSWT. The most conspicuous example is the very broad magnon spectra while the phonon spectra have the linewidth close to the instrumental resolution limit (see Fig. S4~\cite{SM}). The magnon spectra are also much broader than those calculated from the LSWT simulation after being convoluted with the instrumental resolution (\figsref{fig:INS} and \ref{fig:Gamma}). Note that these resolution effects can be precisely calculated by our analysis using TobyFit, which fully takes into account a complex resolution ellipsoid of a real experiment~\cite{YIG_Tobyfit, CuO_Tobyfit, Example_Tobyfit} (See Fig. S4~\cite{SM}). These observations suggest the presence of sizable magnon linewidth broadening or, in other words, magnon decay over the large energy-momentum space. Another noticeable feature is the strong continuum-like signal, whose positions are indicated by the white squares in \figref{fig:INS}a. These additional excitations were observed in the two different INS experiments (\figref{fig:Gamma}a), and cannot be explained by LSWT.

For a full understanding of the observed magnon decay, we analyze the $\q$-dependence of the intrinsic magnon linewidth along D-G and $\textrm{E}_{\textrm{L=0}}$-F in \figref{fig:INS}. The intrinsic half-width at half maximum (HWHM) of the magnon modes (we call it $\Gamma(\q,E_\q)$) is extracted by fitting constant-$\q$ cuts at different $\q$-positions. For the fitting, the calculated magnon modes are convoluted with both the instrumental resolution and the normalized damped simple harmonic oscillator (DSHO) function $f(\q,E)$: 
\begin{equation}
{f(\q,E)} =  \frac{4}{\pi} \frac{\Gamma(\q,E_{\q})\, E\, E_{\q}} {(E^2-{E_{\q}}^2)^2 + 4(\Gamma(\q,E_{\q}) E)^2}, 
\label{eq:DSHO}
\end{equation}
%
where $\q$ and $E$ are the momentum and the energy transfers, respectively. Note that the DSHO model has been widely used to describe quasiparticle decay in a phenomenological way \cite{J.Zhao_DSHO, P.Dai_DSHO, IXS_DSHO}. The fitted results are displayed in \figref{fig:Gamma} as blue solid lines, implying the DSHO model can well describe the data, except for the $\q$-positions where the additional continuum-like excitations appear. \tred{Thus, the fitted results near the continuum-like signal may possess some uncertainty.}
    
\begin{figure}[!t]
\centering
\includegraphics[width=\columnwidth]{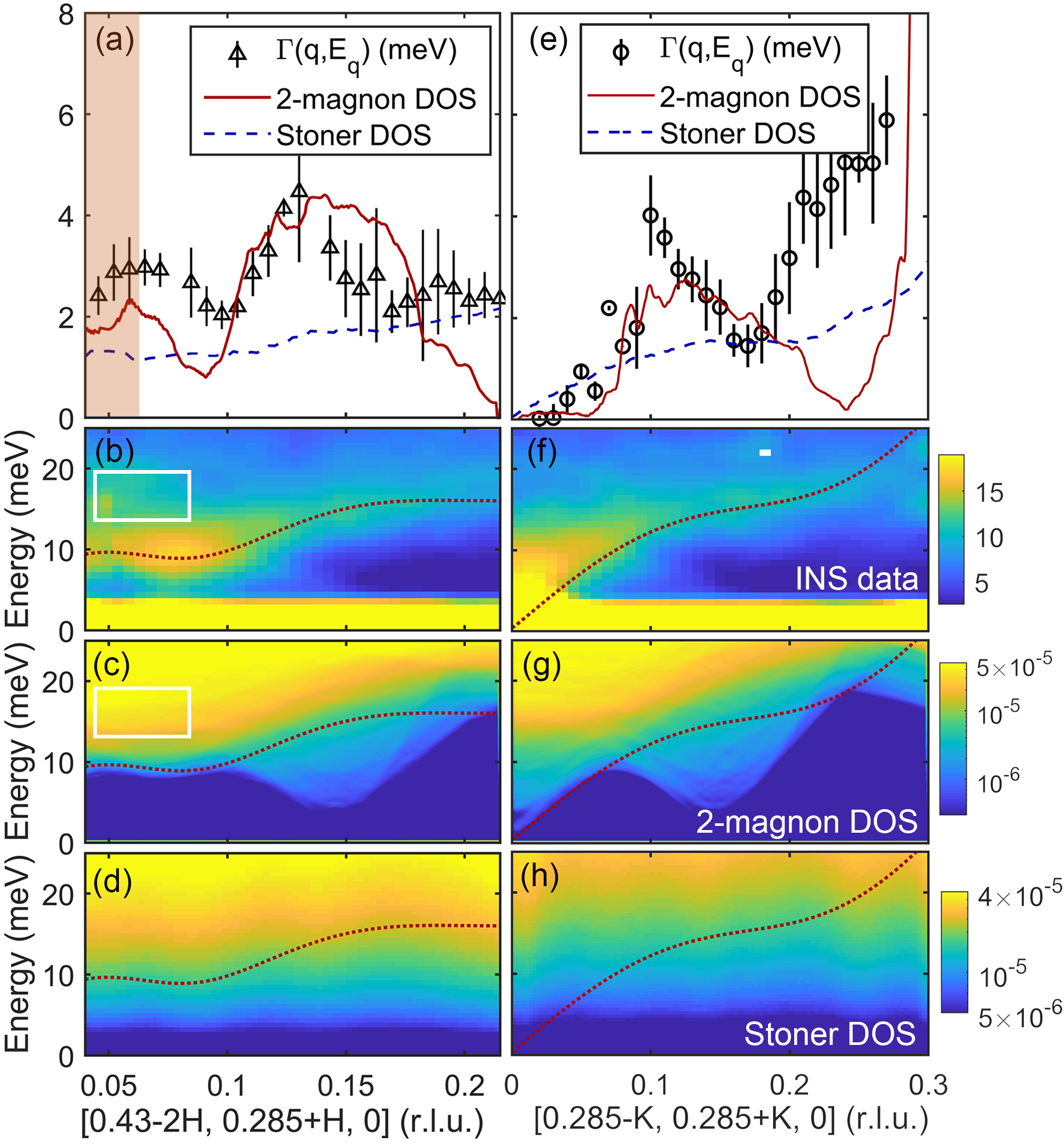}
\caption{(a) Intrinsic HWHM plot of magnon modes ($\Gamma(\q,E_{\q})$) along the [-2H H 0] direction with calculated two-magnon (red solid line) and Stoner (blue dashed line) continuum DOS at $(\q,E_\q)$. $\Gamma(\q,E_\q)$ was extracted from the fitting (\figref{fig:Gamma}) with the instrumental resolution excluded. \tred{Data points in a shaded region may not be reliable due to the overlap with the continuum-like signal.} Error bars indicate the standard deviations of the fitted HWHM. (b) INS data, (c) calculated two-magnon continuum DOS, and (d) calculated Stoner continuum DOS along the same direction as in (a). Red dotted lines indicate the magnon dispersion $E_{\q}$ from the LSWT calculations. White rectangles indicate the region where the continuum-like excitations appear. (e)-(h) are shown along the [-K K 0] direction. For a better presentation, a logarithmic scale was used in (c)-(d) and (g)-(h).
}
\label{fig:DOS}
\end{figure}

\figref{fig:DOS}a and \ref{fig:DOS}e show the extracted $\Gamma(\q,E_{\q})$ along the $\textrm{E}_{\textrm{L=0}}$-F and the D-G lines, which is an important result identifying the origin of the magnon decay. Generally, magnon decay occurs when a magnon branch meets a multi-particle continuum in the $\q$-$E$ space. In \ce{CrB2}, there can be two mechanisms at work for this decay process. One is the decay of a magnon into two magnons under a cubic-order magnon-magnon interaction (the two-magnon continuum), which becomes significant for non-collinear magnetic order~\cite{Chernychev_prl}. Another is the decay of a magnon into an electron-hole pair (the Stoner continuum), which is common in metallic antiferromagnets. The former is known to exhibit unique $\q$-dependence of $\Gamma(\q,E_\q)$ according to the previous studies on insulating TLAFs \cite{YMnO3_nat_comm, Chernychev_prb}, while the latter commonly leads to rather monotonic and straightforward $\q$-dependence of $\Gamma(\q,E_\q)$~\cite{CaFe2As2_gamma, FeMn_gamma, FeGe2_gamma}. In this respect, the magnon decay in \ce{CrB2} is more likely to stem from the two-magnon effect, with the Stoner continuum effect becoming significant at the higher $\q$ points shown in \figref{fig:DOS}.

To further examine $\Gamma(\q,E_\q)$ due to each mechanism specifically for \ce{CrB2}, we calculate the non-interacting two-magnon density of states (DOS) and the Stoner continuum DOS, separately:
\begin{equation}
D(\q,E) =  \frac{1}{N} \sum_{i,j} {\sum_{\k}{\delta (E-E_{\k,i}-E_{\q-\k,j})}},
\label{eq:DOS}
\end{equation}
%
where $\k$ is a set of $\q$-points on the fine, equally spaced mesh in the $1^{\textrm{st}}$ Brillouin zone, $E_{\k,i}$ is the energy dispersion of the $i^{\textrm{th}}$ magnon or electron band in \ce{CrB2}~\cite{SM} and $N$ is a normalization factor. In terms of the magnon decay process, $D(\q$,$E_{\q})$ is a quantity counting the number of possible channels, through which a magnon at ($\q$, $E_{\q}$) can decay into two magnons or an electron-hole pair while satisfying the kinematic constraints of $E_{\q}=E_{\k}-E_{\q-\k}$. Although $D(\q$,$E_{\q})$ is not identical to $\Gamma(\q,E_{\q})$, it is reasonable to compare the two quantities as has been done in previous studies~\cite{LuMno3_PRL_EPG, twomag_DOS, Stoner_DOS}.

\figref{fig:DOS} shows the calculated DOS of each continuum along the two desired $\q$-contours. Except for the high $\q$ region in \figref{fig:DOS}, the two-magnon DOS shows a characteristic behavior in accordance with the experimental $\Gamma(\q,E_{\q})$, while the Stoner continuum DOS shows a simple monotonic behavior and therefore cannot explain the observed $\q$-dependence of $\Gamma(\q,E_\q)$. The latter behavior is observed as the energy of the magnon branches is too low for the Stoner continuum to show its characteristic distribution, given that it is derived from the electron bands with the energy scale of a few \si{\eV}. This result clearly suggests the presence of a two-magnon effect beyond LSWT, which has never been observed in any other metallic magnets yet. \tred{Likewise, the observed continuum-like excitations might as well be from the two-magnon scattering since the Stoner continuum generally does not have INS intensity as comparable to that of magnon modes~\cite{Stoner_intensity}. The sizable two-magnon DOS (\figref{fig:DOS}c) is also consistent with this statement~\cite{LuMno3_PRL_EPG}. For fairer comparison, however, one should derive the exact form of the dynamical structure factor of the two-magnon continuum, which requires more rigorous higher-order calculations~\cite{DynamicalSF_Martin}.}

Such a sizable higher-order effect in \ce{CrB2} can be naturally explained with its non-collinear magnetic structure, which is essential for a non-zero cubic order term in the bosonized spin Hamiltonian. We think that the small magnetic moment of \ce{CrB2} ($\mu_{\mathrm{ordered}}=0.5 \mu_{\mathrm{B}}$) also could be a reason behind a significant higher-order term. However, one has to bear in mind that determining the precise value of $S$ in metallic magnets is problematic due to its fluctuating behavior of itinerant nature. Meanwhile, it is also important to discuss how the contribution of the Stoner continuum to the observed magnon spectra is less dominant in \ce{CrB2}, despite its prevalence in metallic antiferromagnets as a primary factor of magnon damping. One possible reason is that the observed magnon branches are mostly in the low energy region ($< \SI{15}{\meV}$) over the large $\q$-space, different from steep V-shape magnon dispersion seen in other metallic magnets~\cite{Example_Tobyfit, J.Zhao_DSHO, P.Dai_DSHO, FeMn_gamma, FeGe2_gamma, Mn3Sn_npj, Mn3Ge_PRB, Mn3Ge_arxiv}. Since the Stoner continuum DOS is proportional to the energy transfer, those magnon branches may suffer less from the Stoner continuum than those in usual metallic antiferromagnets whereas its influence comes into effect for the high-$\q$ magnon modes with higher energies. It is consistent with the result shown in \figref{fig:DOS}a and \ref{fig:DOS}e; the two-magnon DOS cannot solely explain the experimental $\Gamma(\q,E_{\q})$ at the high-$\q$ region. Another reason is the non-zero ordering wavevector of \ce{CrB2}, enabling its magnon branches to avoid the dense region of the Stoner continuum near the $\Gamma$ point~\cite{SM,Itinerant_Hubbard}. As an example, recent studies on the INS spectra of \ce{Mn3(Sn,Ge)}, which host both steep V-shape dispersion up to \SI{70}{\meV} and a zero or nearby ordering wavevector, have shown severely damped magnon spectra from the Stoner continuum~\cite{Mn3Sn_npj, Mn3Ge_PRB, Mn3Ge_arxiv}.

To summarize, we have studied the magnon spectra of the metallic non-collinear antiferromagnet \ce{CrB2} over the wide $\q$-$E$ space. While the overall magnon dispersion can be explained by LSWT, the data show clear experimental evidence of magnon damping and continuum-like excitations. By analyzing intrinsic linewidth ($\Gamma(\q,E_{\q})$) of the magnon modes, we confirmed characteristic $\q$-dependence of the magnon damping. Further calculations show that it is consistent with $\Gamma(\q,E_{\q})$ due to two-magnon decay while usual Stoner continuum effects become noticeable only at higher $\q$ points. The result reveals, for the first time, the presence of higher-order effects beyond LSWT in metallic antiferromagnets. Our work contributes to a better understanding of magnetism in metallic non-collinear antiferromagnets.

\begin{acknowledgments}
{\it Acknowledgments}:
We thank Martin Mourigal, Pengcheng Dai, and Henrik M. Ronnow for helpful discussions. This work was supported by the Leading Researcher Program of the National Research Foundation of Korea (Grant No. 2020R1A3B2079375) and the Institute for Basic Science in Korea (IBS-R009-G1).
\end{acknowledgments}

\nocite{S1, S2, S3, S4, S5, S6, S7, S8, S9, S10, S11}

\bibliography{CrB2_project_ref}

\begin{thebibliography}{54}%
\makeatletter
\providecommand \@ifxundefined [1]{%
 \@ifx{#1\undefined}
}%
\providecommand \@ifnum [1]{%
 \ifnum #1\expandafter \@firstoftwo
 \else \expandafter \@secondoftwo
 \fi
}%
\providecommand \@ifx [1]{%
 \ifx #1\expandafter \@firstoftwo
 \else \expandafter \@secondoftwo
 \fi
}%
\providecommand \natexlab [1]{#1}%
\providecommand \enquote  [1]{``#1''}%
\providecommand \bibnamefont  [1]{#1}%
\providecommand \bibfnamefont [1]{#1}%
\providecommand \citenamefont [1]{#1}%
\providecommand \href@noop [0]{\@secondoftwo}%
\providecommand \href [0]{\begingroup \@sanitize@url \@href}%
\providecommand \@href[1]{\@@startlink{#1}\@@href}%
\providecommand \@@href[1]{\endgroup#1\@@endlink}%
\providecommand \@sanitize@url [0]{\catcode `\\12\catcode `\$12\catcode
  `\&12\catcode `\#12\catcode `\^12\catcode `\_12\catcode `\%12\relax}%
\providecommand \@@startlink[1]{}%
\providecommand \@@endlink[0]{}%
\providecommand \url  [0]{\begingroup\@sanitize@url \@url }%
\providecommand \@url [1]{\endgroup\@href {#1}{\urlprefix }}%
\providecommand \urlprefix  [0]{URL }%
\providecommand \Eprint [0]{\href }%
\providecommand \doibase [0]{http://dx.doi.org/}%
\providecommand \selectlanguage [0]{\@gobble}%
\providecommand \bibinfo  [0]{\@secondoftwo}%
\providecommand \bibfield  [0]{\@secondoftwo}%
\providecommand \translation [1]{[#1]}%
\providecommand \BibitemOpen [0]{}%
\providecommand \bibitemStop [0]{}%
\providecommand \bibitemNoStop [0]{.\EOS\space}%
\providecommand \EOS [0]{\spacefactor3000\relax}%
\providecommand \BibitemShut  [1]{\csname bibitem#1\endcsname}%
\let\auto@bib@innerbib\@empty
\bibitem [{\citenamefont {Zhitomirsky}\ and\ \citenamefont
  {Chernyshev}(2013)}]{Chernychev_review}%
  \BibitemOpen
  \bibfield  {author} {\bibinfo {author} {\bibfnamefont {M.~E.}\ \bibnamefont
  {Zhitomirsky}}\ and\ \bibinfo {author} {\bibfnamefont {A.~L.}\ \bibnamefont
  {Chernyshev}},\ }\href {\doibase 10.1103/RevModPhys.85.219} {\bibfield
  {journal} {\bibinfo  {journal} {Rev. Mod. Phys.}\ }\textbf {\bibinfo {volume}
  {85}},\ \bibinfo {pages} {219} (\bibinfo {year} {2013})}\BibitemShut
  {NoStop}%
\bibitem [{\citenamefont {Oh}\ \emph {et~al.}(2016)\citenamefont {Oh},
  \citenamefont {Le}, \citenamefont {Nahm}, \citenamefont {Sim}, \citenamefont
  {Jeong}, \citenamefont {Perring}, \citenamefont {Woo}, \citenamefont
  {Nakajima}, \citenamefont {Ohira-Kawamura}, \citenamefont {Yamani},
  \citenamefont {Yoshida}, \citenamefont {Eisaki}, \citenamefont {Cheong},
  \citenamefont {Chernyshev},\ and\ \citenamefont {Park}}]{YMnO3_nat_comm}%
  \BibitemOpen
  \bibfield  {author} {\bibinfo {author} {\bibfnamefont {J.}~\bibnamefont
  {Oh}}, \bibinfo {author} {\bibfnamefont {M.~D.}\ \bibnamefont {Le}}, \bibinfo
  {author} {\bibfnamefont {H.-H.}\ \bibnamefont {Nahm}}, \bibinfo {author}
  {\bibfnamefont {H.}~\bibnamefont {Sim}}, \bibinfo {author} {\bibfnamefont
  {J.}~\bibnamefont {Jeong}}, \bibinfo {author} {\bibfnamefont {T.~G.}\
  \bibnamefont {Perring}}, \bibinfo {author} {\bibfnamefont {H.}~\bibnamefont
  {Woo}}, \bibinfo {author} {\bibfnamefont {K.}~\bibnamefont {Nakajima}},
  \bibinfo {author} {\bibfnamefont {S.}~\bibnamefont {Ohira-Kawamura}},
  \bibinfo {author} {\bibfnamefont {Z.}~\bibnamefont {Yamani}}, \bibinfo
  {author} {\bibfnamefont {Y.}~\bibnamefont {Yoshida}}, \bibinfo {author}
  {\bibfnamefont {H.}~\bibnamefont {Eisaki}}, \bibinfo {author} {\bibfnamefont
  {S.~W.}\ \bibnamefont {Cheong}}, \bibinfo {author} {\bibfnamefont {A.~L.}\
  \bibnamefont {Chernyshev}}, \ and\ \bibinfo {author} {\bibfnamefont {J.-G.}\
  \bibnamefont {Park}},\ }\href {\doibase 10.1038/ncomms13146} {\bibfield
  {journal} {\bibinfo  {journal} {Nat. Commun.}\ }\textbf {\bibinfo {volume}
  {7}},\ \bibinfo {pages} {13146} (\bibinfo {year} {2016})}\BibitemShut
  {NoStop}%
\bibitem [{\citenamefont {Kim}\ \emph {et~al.}(2019)\citenamefont {Kim},
  \citenamefont {Park}, \citenamefont {Leiner},\ and\ \citenamefont
  {Park}}]{TLAF_review}%
  \BibitemOpen
  \bibfield  {author} {\bibinfo {author} {\bibfnamefont {T.}~\bibnamefont
  {Kim}}, \bibinfo {author} {\bibfnamefont {K.}~\bibnamefont {Park}}, \bibinfo
  {author} {\bibfnamefont {J.~C.}\ \bibnamefont {Leiner}}, \ and\ \bibinfo
  {author} {\bibfnamefont {J.-G.}\ \bibnamefont {Park}},\ }\href {\doibase
  10.7566/JPSJ.88.081003} {\bibfield  {journal} {\bibinfo  {journal} {J. Phys.
  Soc. Jpn.}\ }\textbf {\bibinfo {volume} {88}},\ \bibinfo {pages} {081003}
  (\bibinfo {year} {2019})}\BibitemShut {NoStop}%
\bibitem [{\citenamefont {Oh}\ \emph {et~al.}(2013)\citenamefont {Oh},
  \citenamefont {Le}, \citenamefont {Jeong}, \citenamefont {Lee}, \citenamefont
  {Woo}, \citenamefont {Song}, \citenamefont {Perring}, \citenamefont {Buyers},
  \citenamefont {Cheong},\ and\ \citenamefont {Park}}]{LuMno3_PRL_EPG}%
  \BibitemOpen
  \bibfield  {author} {\bibinfo {author} {\bibfnamefont {J.}~\bibnamefont
  {Oh}}, \bibinfo {author} {\bibfnamefont {M.~D.}\ \bibnamefont {Le}}, \bibinfo
  {author} {\bibfnamefont {J.}~\bibnamefont {Jeong}}, \bibinfo {author}
  {\bibfnamefont {J.-h.}\ \bibnamefont {Lee}}, \bibinfo {author} {\bibfnamefont
  {H.}~\bibnamefont {Woo}}, \bibinfo {author} {\bibfnamefont {W.-Y.}\
  \bibnamefont {Song}}, \bibinfo {author} {\bibfnamefont {T.~G.}\ \bibnamefont
  {Perring}}, \bibinfo {author} {\bibfnamefont {W.~J.~L.}\ \bibnamefont
  {Buyers}}, \bibinfo {author} {\bibfnamefont {S.~W.}\ \bibnamefont {Cheong}},
  \ and\ \bibinfo {author} {\bibfnamefont {J.-G.}\ \bibnamefont {Park}},\
  }\href {\doibase 10.1103/PhysRevLett.111.257202} {\bibfield  {journal}
  {\bibinfo  {journal} {Phys. Rev. Lett.}\ }\textbf {\bibinfo {volume} {111}},\
  \bibinfo {pages} {257202} (\bibinfo {year} {2013})}\BibitemShut {NoStop}%
\bibitem [{\citenamefont {Petit}\ \emph {et~al.}(2007)\citenamefont {Petit},
  \citenamefont {Moussa}, \citenamefont {Hennion}, \citenamefont {Pailhès},
  \citenamefont {Pinsard-Gaudart},\ and\ \citenamefont
  {Ivanov}}]{MP_coupling_Petit}%
  \BibitemOpen
  \bibfield  {author} {\bibinfo {author} {\bibfnamefont {S.}~\bibnamefont
  {Petit}}, \bibinfo {author} {\bibfnamefont {F.}~\bibnamefont {Moussa}},
  \bibinfo {author} {\bibfnamefont {M.}~\bibnamefont {Hennion}}, \bibinfo
  {author} {\bibfnamefont {S.}~\bibnamefont {Pailhès}}, \bibinfo {author}
  {\bibfnamefont {L.}~\bibnamefont {Pinsard-Gaudart}}, \ and\ \bibinfo {author}
  {\bibfnamefont {A.}~\bibnamefont {Ivanov}},\ }\href {\doibase
  10.1103/PhysRevLett.99.266604} {\bibfield  {journal} {\bibinfo  {journal}
  {Phys. Rev. Lett.}\ }\textbf {\bibinfo {volume} {99}},\ \bibinfo {pages}
  {266604} (\bibinfo {year} {2007})}\BibitemShut {NoStop}%
\bibitem [{\citenamefont {Kim}\ \emph {et~al.}(2018)\citenamefont {Kim},
  \citenamefont {Leiner}, \citenamefont {Park}, \citenamefont {Oh},
  \citenamefont {Sim}, \citenamefont {Iida}, \citenamefont {Kamazawa},\ and\
  \citenamefont {Park}}]{HoMnO3_EPG}%
  \BibitemOpen
  \bibfield  {author} {\bibinfo {author} {\bibfnamefont {T.}~\bibnamefont
  {Kim}}, \bibinfo {author} {\bibfnamefont {J.~C.}\ \bibnamefont {Leiner}},
  \bibinfo {author} {\bibfnamefont {K.}~\bibnamefont {Park}}, \bibinfo {author}
  {\bibfnamefont {J.}~\bibnamefont {Oh}}, \bibinfo {author} {\bibfnamefont
  {H.}~\bibnamefont {Sim}}, \bibinfo {author} {\bibfnamefont {K.}~\bibnamefont
  {Iida}}, \bibinfo {author} {\bibfnamefont {K.}~\bibnamefont {Kamazawa}}, \
  and\ \bibinfo {author} {\bibfnamefont {J.-G.}\ \bibnamefont {Park}},\ }\href
  {\doibase 10.1103/PhysRevB.97.201113} {\bibfield  {journal} {\bibinfo
  {journal} {Phys. Rev. B}\ }\textbf {\bibinfo {volume} {97}},\ \bibinfo
  {pages} {201113} (\bibinfo {year} {2018})}\BibitemShut {NoStop}%
\bibitem [{\citenamefont {Park}\ \emph {et~al.}(2016)\citenamefont {Park},
  \citenamefont {Oh}, \citenamefont {Leiner}, \citenamefont {Jeong},
  \citenamefont {Rule}, \citenamefont {Le},\ and\ \citenamefont
  {Park}}]{CuCrO2_PRB}%
  \BibitemOpen
  \bibfield  {author} {\bibinfo {author} {\bibfnamefont {K.}~\bibnamefont
  {Park}}, \bibinfo {author} {\bibfnamefont {J.}~\bibnamefont {Oh}}, \bibinfo
  {author} {\bibfnamefont {J.~C.}\ \bibnamefont {Leiner}}, \bibinfo {author}
  {\bibfnamefont {J.}~\bibnamefont {Jeong}}, \bibinfo {author} {\bibfnamefont
  {K.~C.}\ \bibnamefont {Rule}}, \bibinfo {author} {\bibfnamefont {M.~D.}\
  \bibnamefont {Le}}, \ and\ \bibinfo {author} {\bibfnamefont {J.-G.}\
  \bibnamefont {Park}},\ }\href {\doibase 10.1103/PhysRevB.94.104421}
  {\bibfield  {journal} {\bibinfo  {journal} {Phys. Rev. B}\ }\textbf {\bibinfo
  {volume} {94}},\ \bibinfo {pages} {104421} (\bibinfo {year}
  {2016})}\BibitemShut {NoStop}%
\bibitem [{\citenamefont {Tóth}\ \emph {et~al.}(2016)\citenamefont {Tóth},
  \citenamefont {Wehinger}, \citenamefont {Rolfs}, \citenamefont {Birol},
  \citenamefont {Stuhr}, \citenamefont {Takatsu}, \citenamefont {Kimura},
  \citenamefont {Kimura}, \citenamefont {Rønnow},\ and\ \citenamefont
  {Rüegg}}]{LiCrO2_Toth}%
  \BibitemOpen
  \bibfield  {author} {\bibinfo {author} {\bibfnamefont {S.}~\bibnamefont
  {Tóth}}, \bibinfo {author} {\bibfnamefont {B.}~\bibnamefont {Wehinger}},
  \bibinfo {author} {\bibfnamefont {K.}~\bibnamefont {Rolfs}}, \bibinfo
  {author} {\bibfnamefont {T.}~\bibnamefont {Birol}}, \bibinfo {author}
  {\bibfnamefont {U.}~\bibnamefont {Stuhr}}, \bibinfo {author} {\bibfnamefont
  {H.}~\bibnamefont {Takatsu}}, \bibinfo {author} {\bibfnamefont
  {K.}~\bibnamefont {Kimura}}, \bibinfo {author} {\bibfnamefont
  {T.}~\bibnamefont {Kimura}}, \bibinfo {author} {\bibfnamefont {H.~M.}\
  \bibnamefont {Rønnow}}, \ and\ \bibinfo {author} {\bibfnamefont
  {C.}~\bibnamefont {Rüegg}},\ }\href {\doibase 10.1038/ncomms13547}
  {\bibfield  {journal} {\bibinfo  {journal} {Nat. Commun.}\ }\textbf {\bibinfo
  {volume} {7}},\ \bibinfo {pages} {13547} (\bibinfo {year}
  {2016})}\BibitemShut {NoStop}%
\bibitem [{\citenamefont {Ma}\ \emph {et~al.}(2016)\citenamefont {Ma},
  \citenamefont {Kamiya}, \citenamefont {Hong}, \citenamefont {Cao},
  \citenamefont {Ehlers}, \citenamefont {Tian}, \citenamefont {Batista},
  \citenamefont {Dun}, \citenamefont {Zhou},\ and\ \citenamefont
  {Matsuda}}]{BCSO_J.Ma}%
  \BibitemOpen
  \bibfield  {author} {\bibinfo {author} {\bibfnamefont {J.}~\bibnamefont
  {Ma}}, \bibinfo {author} {\bibfnamefont {Y.}~\bibnamefont {Kamiya}}, \bibinfo
  {author} {\bibfnamefont {T.}~\bibnamefont {Hong}}, \bibinfo {author}
  {\bibfnamefont {H.}~\bibnamefont {Cao}}, \bibinfo {author} {\bibfnamefont
  {G.}~\bibnamefont {Ehlers}}, \bibinfo {author} {\bibfnamefont
  {W.}~\bibnamefont {Tian}}, \bibinfo {author} {\bibfnamefont {C.}~\bibnamefont
  {Batista}}, \bibinfo {author} {\bibfnamefont {Z.}~\bibnamefont {Dun}},
  \bibinfo {author} {\bibfnamefont {H.}~\bibnamefont {Zhou}}, \ and\ \bibinfo
  {author} {\bibfnamefont {M.}~\bibnamefont {Matsuda}},\ }\href {\doibase
  10.1103/PhysRevLett.116.087201} {\bibfield  {journal} {\bibinfo  {journal}
  {Phys. Rev. Lett.}\ }\textbf {\bibinfo {volume} {116}},\ \bibinfo {pages}
  {087201} (\bibinfo {year} {2016})}\BibitemShut {NoStop}%
\bibitem [{\citenamefont {Verresen}\ \emph {et~al.}(2019)\citenamefont
  {Verresen}, \citenamefont {Moessner},\ and\ \citenamefont
  {Pollmann}}]{BCSO_nat_phys}%
  \BibitemOpen
  \bibfield  {author} {\bibinfo {author} {\bibfnamefont {R.}~\bibnamefont
  {Verresen}}, \bibinfo {author} {\bibfnamefont {R.}~\bibnamefont {Moessner}},
  \ and\ \bibinfo {author} {\bibfnamefont {F.}~\bibnamefont {Pollmann}},\
  }\href {\doibase 10.1038/s41567-019-0535-3} {\bibfield  {journal} {\bibinfo
  {journal} {Nat. Phys.}\ }\textbf {\bibinfo {volume} {15}},\ \bibinfo {pages}
  {750} (\bibinfo {year} {2019})}\BibitemShut {NoStop}%
\bibitem [{\citenamefont {Prange}\ and\ \citenamefont
  {Korenman}(1979)}]{Itinerant_Heisenberg}%
  \BibitemOpen
  \bibfield  {author} {\bibinfo {author} {\bibfnamefont {R.~E.}\ \bibnamefont
  {Prange}}\ and\ \bibinfo {author} {\bibfnamefont {V.}~\bibnamefont
  {Korenman}},\ }\href {\doibase 10.1103/PhysRevB.19.4691} {\bibfield
  {journal} {\bibinfo  {journal} {Phys. Rev. B}\ }\textbf {\bibinfo {volume}
  {19}},\ \bibinfo {pages} {4691} (\bibinfo {year} {1979})}\BibitemShut
  {NoStop}%
\bibitem [{\citenamefont {Nakatsuji}\ \emph {et~al.}(2015)\citenamefont
  {Nakatsuji}, \citenamefont {Kiyohara},\ and\ \citenamefont
  {Higo}}]{Mn3Sn_AHE}%
  \BibitemOpen
  \bibfield  {author} {\bibinfo {author} {\bibfnamefont {S.}~\bibnamefont
  {Nakatsuji}}, \bibinfo {author} {\bibfnamefont {N.}~\bibnamefont {Kiyohara}},
  \ and\ \bibinfo {author} {\bibfnamefont {T.}~\bibnamefont {Higo}},\ }\href
  {\doibase 10.1038/nature15723} {\bibfield  {journal} {\bibinfo  {journal}
  {Nature}\ }\textbf {\bibinfo {volume} {527}},\ \bibinfo {pages} {212}
  (\bibinfo {year} {2015})}\BibitemShut {NoStop}%
\bibitem [{\citenamefont {Nayak}\ \emph {et~al.}(2016)\citenamefont {Nayak},
  \citenamefont {Fischer}, \citenamefont {Sun}, \citenamefont {Yan},
  \citenamefont {Karel}, \citenamefont {Komarek}, \citenamefont {Shekhar},
  \citenamefont {Kumar}, \citenamefont {Schnelle}, \citenamefont {Kübler},
  \citenamefont {Felser},\ and\ \citenamefont {Parkin}}]{Mn3Ge_AHE}%
  \BibitemOpen
  \bibfield  {author} {\bibinfo {author} {\bibfnamefont {A.~K.}\ \bibnamefont
  {Nayak}}, \bibinfo {author} {\bibfnamefont {J.~E.}\ \bibnamefont {Fischer}},
  \bibinfo {author} {\bibfnamefont {Y.}~\bibnamefont {Sun}}, \bibinfo {author}
  {\bibfnamefont {B.}~\bibnamefont {Yan}}, \bibinfo {author} {\bibfnamefont
  {J.}~\bibnamefont {Karel}}, \bibinfo {author} {\bibfnamefont {A.~C.}\
  \bibnamefont {Komarek}}, \bibinfo {author} {\bibfnamefont {C.}~\bibnamefont
  {Shekhar}}, \bibinfo {author} {\bibfnamefont {N.}~\bibnamefont {Kumar}},
  \bibinfo {author} {\bibfnamefont {W.}~\bibnamefont {Schnelle}}, \bibinfo
  {author} {\bibfnamefont {J.}~\bibnamefont {Kübler}}, \bibinfo {author}
  {\bibfnamefont {C.}~\bibnamefont {Felser}}, \ and\ \bibinfo {author}
  {\bibfnamefont {S.~S.~P.}\ \bibnamefont {Parkin}},\ }\href {\doibase
  10.1126/sciadv.1501870} {\bibfield  {journal} {\bibinfo  {journal} {Sci.
  Adv.}\ }\textbf {\bibinfo {volume} {2}},\ \bibinfo {pages} {e1501870}
  (\bibinfo {year} {2016})}\BibitemShut {NoStop}%
\bibitem [{\citenamefont {Kübler}\ and\ \citenamefont
  {Felser}(2014)}]{AHE_Kubler}%
  \BibitemOpen
  \bibfield  {author} {\bibinfo {author} {\bibfnamefont {J.}~\bibnamefont
  {Kübler}}\ and\ \bibinfo {author} {\bibfnamefont {C.}~\bibnamefont
  {Felser}},\ }\href {\doibase 10.1209/0295-5075/108/67001} {\bibfield
  {journal} {\bibinfo  {journal} {EPL (Europhys. Lett.)}\ }\textbf {\bibinfo
  {volume} {108}},\ \bibinfo {pages} {67001} (\bibinfo {year}
  {2014})}\BibitemShut {NoStop}%
\bibitem [{\citenamefont {Park}\ \emph {et~al.}(2018)\citenamefont {Park},
  \citenamefont {Oh}, \citenamefont {Uhlířová}, \citenamefont {Jackson},
  \citenamefont {Deák}, \citenamefont {Szunyogh}, \citenamefont {Lee},
  \citenamefont {Cho}, \citenamefont {Kim}, \citenamefont {Walker},
  \citenamefont {Adroja}, \citenamefont {Sechovský},\ and\ \citenamefont
  {Park}}]{Mn3Sn_npj}%
  \BibitemOpen
  \bibfield  {author} {\bibinfo {author} {\bibfnamefont {P.}~\bibnamefont
  {Park}}, \bibinfo {author} {\bibfnamefont {J.}~\bibnamefont {Oh}}, \bibinfo
  {author} {\bibfnamefont {K.}~\bibnamefont {Uhlířová}}, \bibinfo {author}
  {\bibfnamefont {J.}~\bibnamefont {Jackson}}, \bibinfo {author} {\bibfnamefont
  {A.}~\bibnamefont {Deák}}, \bibinfo {author} {\bibfnamefont
  {L.}~\bibnamefont {Szunyogh}}, \bibinfo {author} {\bibfnamefont {K.~H.}\
  \bibnamefont {Lee}}, \bibinfo {author} {\bibfnamefont {H.}~\bibnamefont
  {Cho}}, \bibinfo {author} {\bibfnamefont {H.-L.}\ \bibnamefont {Kim}},
  \bibinfo {author} {\bibfnamefont {H.~C.}\ \bibnamefont {Walker}}, \bibinfo
  {author} {\bibfnamefont {D.}~\bibnamefont {Adroja}}, \bibinfo {author}
  {\bibfnamefont {V.}~\bibnamefont {Sechovský}}, \ and\ \bibinfo {author}
  {\bibfnamefont {J.-G.}\ \bibnamefont {Park}},\ }\href {\doibase
  10.1038/s41535-018-0137-9} {\bibfield  {journal} {\bibinfo  {journal} {npj
  Quantum Mater.}\ }\textbf {\bibinfo {volume} {3}},\ \bibinfo {pages} {63}
  (\bibinfo {year} {2018})}\BibitemShut {NoStop}%
\bibitem [{\citenamefont {Funahashi}\ \emph {et~al.}(1977)\citenamefont
  {Funahashi}, \citenamefont {Hamaguchi}, \citenamefont {Tanaka},\ and\
  \citenamefont {Bannai}}]{NDold_CrB2}%
  \BibitemOpen
  \bibfield  {author} {\bibinfo {author} {\bibfnamefont {S.}~\bibnamefont
  {Funahashi}}, \bibinfo {author} {\bibfnamefont {Y.}~\bibnamefont
  {Hamaguchi}}, \bibinfo {author} {\bibfnamefont {T.}~\bibnamefont {Tanaka}}, \
  and\ \bibinfo {author} {\bibfnamefont {E.}~\bibnamefont {Bannai}},\ }\href
  {\doibase https://doi.org/10.1016/0038-1098(77)90969-3} {\bibfield  {journal}
  {\bibinfo  {journal} {Solid State Communications}\ }\textbf {\bibinfo
  {volume} {23}},\ \bibinfo {pages} {859} (\bibinfo {year} {1977})}\BibitemShut
  {NoStop}%
\bibitem [{\citenamefont {Kaya}\ \emph {et~al.}(2009)\citenamefont {Kaya},
  \citenamefont {Kousaka}, \citenamefont {Kakurai}, \citenamefont {Takeda},\
  and\ \citenamefont {Akimitsu}}]{NDnew_CrB2}%
  \BibitemOpen
  \bibfield  {author} {\bibinfo {author} {\bibfnamefont {E.}~\bibnamefont
  {Kaya}}, \bibinfo {author} {\bibfnamefont {Y.}~\bibnamefont {Kousaka}},
  \bibinfo {author} {\bibfnamefont {K.}~\bibnamefont {Kakurai}}, \bibinfo
  {author} {\bibfnamefont {M.}~\bibnamefont {Takeda}}, \ and\ \bibinfo {author}
  {\bibfnamefont {J.}~\bibnamefont {Akimitsu}},\ }\href {\doibase
  https://doi.org/10.1016/j.physb.2009.06.010} {\bibfield  {journal} {\bibinfo
  {journal} {Physica B: Condensed Matter}\ }\textbf {\bibinfo {volume} {404}},\
  \bibinfo {pages} {2524} (\bibinfo {year} {2009})}\BibitemShut {NoStop}%
\bibitem [{\citenamefont {Ivanov}\ and\ \citenamefont
  {Richter}(1995)}]{TLAF_phase_diagram}%
  \BibitemOpen
  \bibfield  {author} {\bibinfo {author} {\bibfnamefont {N.~B.}\ \bibnamefont
  {Ivanov}}\ and\ \bibinfo {author} {\bibfnamefont {J.}~\bibnamefont
  {Richter}},\ }\href {\doibase https://doi.org/10.1016/0304-8853(94)00530-3}
  {\bibfield  {journal} {\bibinfo  {journal} {J. Magn. Magn. Mater.}\ }\textbf
  {\bibinfo {volume} {140-144}},\ \bibinfo {pages} {1965} (\bibinfo {year}
  {1995})}\BibitemShut {NoStop}%
\bibitem [{\citenamefont {Bauer}\ \emph {et~al.}(2014)\citenamefont {Bauer},
  \citenamefont {Regnat}, \citenamefont {Blum}, \citenamefont
  {Gottlieb-Schönmeyer}, \citenamefont {Pedersen}, \citenamefont {Meven},
  \citenamefont {Wurmehl}, \citenamefont {Kuneš},\ and\ \citenamefont
  {Pfleiderer}}]{Bauer_CrB2}%
  \BibitemOpen
  \bibfield  {author} {\bibinfo {author} {\bibfnamefont {A.}~\bibnamefont
  {Bauer}}, \bibinfo {author} {\bibfnamefont {A.}~\bibnamefont {Regnat}},
  \bibinfo {author} {\bibfnamefont {C.~G.~F.}\ \bibnamefont {Blum}}, \bibinfo
  {author} {\bibfnamefont {S.}~\bibnamefont {Gottlieb-Schönmeyer}}, \bibinfo
  {author} {\bibfnamefont {B.}~\bibnamefont {Pedersen}}, \bibinfo {author}
  {\bibfnamefont {M.}~\bibnamefont {Meven}}, \bibinfo {author} {\bibfnamefont
  {S.}~\bibnamefont {Wurmehl}}, \bibinfo {author} {\bibfnamefont
  {J.}~\bibnamefont {Kuneš}}, \ and\ \bibinfo {author} {\bibfnamefont
  {C.}~\bibnamefont {Pfleiderer}},\ }\href {\doibase
  10.1103/PhysRevB.90.064414} {\bibfield  {journal} {\bibinfo  {journal}
  {Physical Review B}\ }\textbf {\bibinfo {volume} {90}},\ \bibinfo {pages}
  {064414} (\bibinfo {year} {2014})}\BibitemShut {NoStop}%
\bibitem [{SM()}]{SM}%
  \BibitemOpen
  \href@noop {} {}\bibinfo {note} {See Supplemental Material at [URL] for
  detailed sample preparation, the phonon spectra of \ce{CrB2}, the
  significance of long-range RKKY interaction, detailed instrumental resolution
  issues, the electron band calculation with more discussions about the Stoner
  continuum, and the reciprocal lattice coordinates of the $\q$-points used in
  the main text.}\BibitemShut {Stop}%
\bibitem [{\citenamefont {Castaing}\ \emph {et~al.}(1972)\citenamefont
  {Castaing}, \citenamefont {Costa}, \citenamefont {Heritier},\ and\
  \citenamefont {Lederer}}]{MT_CrB2}%
  \BibitemOpen
  \bibfield  {author} {\bibinfo {author} {\bibfnamefont {J.}~\bibnamefont
  {Castaing}}, \bibinfo {author} {\bibfnamefont {P.}~\bibnamefont {Costa}},
  \bibinfo {author} {\bibfnamefont {M.}~\bibnamefont {Heritier}}, \ and\
  \bibinfo {author} {\bibfnamefont {P.}~\bibnamefont {Lederer}},\ }\href
  {\doibase https://doi.org/10.1016/0022-3697(72)90035-2} {\bibfield  {journal}
  {\bibinfo  {journal} {Journal of Physics and Chemistry of Solids}\ }\textbf
  {\bibinfo {volume} {33}},\ \bibinfo {pages} {533} (\bibinfo {year}
  {1972})}\BibitemShut {NoStop}%
\bibitem [{\citenamefont {Ewings}\ \emph {et~al.}(2019)\citenamefont {Ewings},
  \citenamefont {Stewart}, \citenamefont {Perring}, \citenamefont {Bewley},
  \citenamefont {Le}, \citenamefont {Raspino}, \citenamefont {Pooley},
  \citenamefont {Škoro}, \citenamefont {Waller}, \citenamefont {Zacek},
  \citenamefont {Smith},\ and\ \citenamefont {Riehl-Shaw}}]{MAPS_ref}%
  \BibitemOpen
  \bibfield  {author} {\bibinfo {author} {\bibfnamefont {R.~A.}\ \bibnamefont
  {Ewings}}, \bibinfo {author} {\bibfnamefont {J.~R.}\ \bibnamefont {Stewart}},
  \bibinfo {author} {\bibfnamefont {T.~G.}\ \bibnamefont {Perring}}, \bibinfo
  {author} {\bibfnamefont {R.~I.}\ \bibnamefont {Bewley}}, \bibinfo {author}
  {\bibfnamefont {M.~D.}\ \bibnamefont {Le}}, \bibinfo {author} {\bibfnamefont
  {D.}~\bibnamefont {Raspino}}, \bibinfo {author} {\bibfnamefont {D.~E.}\
  \bibnamefont {Pooley}}, \bibinfo {author} {\bibfnamefont {G.}~\bibnamefont
  {Škoro}}, \bibinfo {author} {\bibfnamefont {S.~P.}\ \bibnamefont {Waller}},
  \bibinfo {author} {\bibfnamefont {D.}~\bibnamefont {Zacek}}, \bibinfo
  {author} {\bibfnamefont {C.~A.}\ \bibnamefont {Smith}}, \ and\ \bibinfo
  {author} {\bibfnamefont {R.~C.}\ \bibnamefont {Riehl-Shaw}},\ }\href
  {\doibase 10.1063/1.5086255} {\bibfield  {journal} {\bibinfo  {journal} {Rev.
  Sci. Instrum.}\ }\textbf {\bibinfo {volume} {90}},\ \bibinfo {pages} {035110}
  (\bibinfo {year} {2019})}\BibitemShut {NoStop}%
\bibitem [{MAP()}]{MAPS_data_number}%
  \BibitemOpen
  \href@noop {} {}\bibinfo {note} {Park, J.-G. et al. 1810057, STFC ISIS
  Neutron and Muon Source, https://doi.org/10.5286/ISIS.E.98000070
  (2018)}\BibitemShut {NoStop}%
\bibitem [{\citenamefont {Ewings}\ \emph {et~al.}(2016)\citenamefont {Ewings},
  \citenamefont {Buts}, \citenamefont {Le}, \citenamefont {van Duijn},
  \citenamefont {Bustinduy},\ and\ \citenamefont {Perring}}]{Horace_ref}%
  \BibitemOpen
  \bibfield  {author} {\bibinfo {author} {\bibfnamefont {R.~A.}\ \bibnamefont
  {Ewings}}, \bibinfo {author} {\bibfnamefont {A.}~\bibnamefont {Buts}},
  \bibinfo {author} {\bibfnamefont {M.~D.}\ \bibnamefont {Le}}, \bibinfo
  {author} {\bibfnamefont {J.}~\bibnamefont {van Duijn}}, \bibinfo {author}
  {\bibfnamefont {I.}~\bibnamefont {Bustinduy}}, \ and\ \bibinfo {author}
  {\bibfnamefont {T.~G.}\ \bibnamefont {Perring}},\ }\href {\doibase
  https://doi.org/10.1016/j.nima.2016.07.036} {\bibfield  {journal} {\bibinfo
  {journal} {Nuclear Instruments and Methods in Physics Research Section A:
  Accelerators, Spectrometers, Detectors and Associated Equipment}\ }\textbf
  {\bibinfo {volume} {834}},\ \bibinfo {pages} {132} (\bibinfo {year}
  {2016})}\BibitemShut {NoStop}%
\bibitem [{\citenamefont {Toth}\ and\ \citenamefont {Lake}(2015)}]{SpinW_ref}%
  \BibitemOpen
  \bibfield  {author} {\bibinfo {author} {\bibfnamefont {S.}~\bibnamefont
  {Toth}}\ and\ \bibinfo {author} {\bibfnamefont {B.}~\bibnamefont {Lake}},\
  }\href {\doibase 10.1088/0953-8984/27/16/166002} {\bibfield  {journal}
  {\bibinfo  {journal} {J. Phys.: Condens. Matter}\ }\textbf {\bibinfo {volume}
  {27}},\ \bibinfo {pages} {166002} (\bibinfo {year} {2015})}\BibitemShut
  {NoStop}%
\bibitem [{\citenamefont {Princep}\ \emph {et~al.}(2017)\citenamefont
  {Princep}, \citenamefont {Ewings}, \citenamefont {Ward}, \citenamefont
  {Tóth}, \citenamefont {Dubs}, \citenamefont {Prabhakaran},\ and\
  \citenamefont {Boothroyd}}]{YIG_Tobyfit}%
  \BibitemOpen
  \bibfield  {author} {\bibinfo {author} {\bibfnamefont {A.~J.}\ \bibnamefont
  {Princep}}, \bibinfo {author} {\bibfnamefont {R.~A.}\ \bibnamefont {Ewings}},
  \bibinfo {author} {\bibfnamefont {S.}~\bibnamefont {Ward}}, \bibinfo {author}
  {\bibfnamefont {S.}~\bibnamefont {Tóth}}, \bibinfo {author} {\bibfnamefont
  {C.}~\bibnamefont {Dubs}}, \bibinfo {author} {\bibfnamefont {D.}~\bibnamefont
  {Prabhakaran}}, \ and\ \bibinfo {author} {\bibfnamefont {A.~T.}\ \bibnamefont
  {Boothroyd}},\ }\href {\doibase 10.1038/s41535-017-0067-y} {\bibfield
  {journal} {\bibinfo  {journal} {npj Quantum Mater.}\ }\textbf {\bibinfo
  {volume} {2}},\ \bibinfo {pages} {63} (\bibinfo {year} {2017})}\BibitemShut
  {NoStop}%
\bibitem [{\citenamefont {Jacobsen}\ \emph {et~al.}(2018)\citenamefont
  {Jacobsen}, \citenamefont {Gaw}, \citenamefont {Princep}, \citenamefont
  {Hamilton}, \citenamefont {Tóth}, \citenamefont {Ewings}, \citenamefont
  {Enderle}, \citenamefont {Wheeler}, \citenamefont {Prabhakaran},\ and\
  \citenamefont {Boothroyd}}]{CuO_Tobyfit}%
  \BibitemOpen
  \bibfield  {author} {\bibinfo {author} {\bibfnamefont {H.}~\bibnamefont
  {Jacobsen}}, \bibinfo {author} {\bibfnamefont {S.~M.}\ \bibnamefont {Gaw}},
  \bibinfo {author} {\bibfnamefont {A.~J.}\ \bibnamefont {Princep}}, \bibinfo
  {author} {\bibfnamefont {E.}~\bibnamefont {Hamilton}}, \bibinfo {author}
  {\bibfnamefont {S.}~\bibnamefont {Tóth}}, \bibinfo {author} {\bibfnamefont
  {R.~A.}\ \bibnamefont {Ewings}}, \bibinfo {author} {\bibfnamefont
  {M.}~\bibnamefont {Enderle}}, \bibinfo {author} {\bibfnamefont {E.~M.~H.}\
  \bibnamefont {Wheeler}}, \bibinfo {author} {\bibfnamefont {D.}~\bibnamefont
  {Prabhakaran}}, \ and\ \bibinfo {author} {\bibfnamefont {A.~T.}\ \bibnamefont
  {Boothroyd}},\ }\href {\doibase 10.1103/PhysRevB.97.144401} {\bibfield
  {journal} {\bibinfo  {journal} {Phys. Rev. B}\ }\textbf {\bibinfo {volume}
  {97}},\ \bibinfo {pages} {144401} (\bibinfo {year} {2018})}\BibitemShut
  {NoStop}%
\bibitem [{\citenamefont {Wang}\ \emph {et~al.}(2015)\citenamefont {Wang},
  \citenamefont {Valdivia}, \citenamefont {Yi}, \citenamefont {Chen},
  \citenamefont {Zhang}, \citenamefont {Ewings}, \citenamefont {Perring},
  \citenamefont {Zhao}, \citenamefont {Harriger}, \citenamefont {Lynn},
  \citenamefont {Bourret-Courchesne}, \citenamefont {Dai}, \citenamefont {Lee},
  \citenamefont {Yao},\ and\ \citenamefont {Birgeneau}}]{Example_Tobyfit}%
  \BibitemOpen
  \bibfield  {author} {\bibinfo {author} {\bibfnamefont {M.}~\bibnamefont
  {Wang}}, \bibinfo {author} {\bibfnamefont {P.}~\bibnamefont {Valdivia}},
  \bibinfo {author} {\bibfnamefont {M.}~\bibnamefont {Yi}}, \bibinfo {author}
  {\bibfnamefont {J.~X.}\ \bibnamefont {Chen}}, \bibinfo {author}
  {\bibfnamefont {W.~L.}\ \bibnamefont {Zhang}}, \bibinfo {author}
  {\bibfnamefont {R.~A.}\ \bibnamefont {Ewings}}, \bibinfo {author}
  {\bibfnamefont {T.~G.}\ \bibnamefont {Perring}}, \bibinfo {author}
  {\bibfnamefont {Y.}~\bibnamefont {Zhao}}, \bibinfo {author} {\bibfnamefont
  {L.~W.}\ \bibnamefont {Harriger}}, \bibinfo {author} {\bibfnamefont {J.~W.}\
  \bibnamefont {Lynn}}, \bibinfo {author} {\bibfnamefont {E.}~\bibnamefont
  {Bourret-Courchesne}}, \bibinfo {author} {\bibfnamefont {P.}~\bibnamefont
  {Dai}}, \bibinfo {author} {\bibfnamefont {D.~H.}\ \bibnamefont {Lee}},
  \bibinfo {author} {\bibfnamefont {D.~X.}\ \bibnamefont {Yao}}, \ and\
  \bibinfo {author} {\bibfnamefont {R.~J.}\ \bibnamefont {Birgeneau}},\ }\href
  {\doibase 10.1103/PhysRevB.92.041109} {\bibfield  {journal} {\bibinfo
  {journal} {Phys. Rev. B}\ }\textbf {\bibinfo {volume} {92}},\ \bibinfo
  {pages} {041109} (\bibinfo {year} {2015})}\BibitemShut {NoStop}%
\bibitem [{\citenamefont {Zhao}\ \emph {et~al.}(2009)\citenamefont {Zhao},
  \citenamefont {Adroja}, \citenamefont {Yao}, \citenamefont {Bewley},
  \citenamefont {Li}, \citenamefont {Wang}, \citenamefont {Wu}, \citenamefont
  {Chen}, \citenamefont {Hu},\ and\ \citenamefont {Dai}}]{J.Zhao_DSHO}%
  \BibitemOpen
  \bibfield  {author} {\bibinfo {author} {\bibfnamefont {J.}~\bibnamefont
  {Zhao}}, \bibinfo {author} {\bibfnamefont {D.~T.}\ \bibnamefont {Adroja}},
  \bibinfo {author} {\bibfnamefont {D.-X.}\ \bibnamefont {Yao}}, \bibinfo
  {author} {\bibfnamefont {R.}~\bibnamefont {Bewley}}, \bibinfo {author}
  {\bibfnamefont {S.}~\bibnamefont {Li}}, \bibinfo {author} {\bibfnamefont
  {X.~F.}\ \bibnamefont {Wang}}, \bibinfo {author} {\bibfnamefont
  {G.}~\bibnamefont {Wu}}, \bibinfo {author} {\bibfnamefont {X.~H.}\
  \bibnamefont {Chen}}, \bibinfo {author} {\bibfnamefont {J.}~\bibnamefont
  {Hu}}, \ and\ \bibinfo {author} {\bibfnamefont {P.}~\bibnamefont {Dai}},\
  }\href {\doibase 10.1038/nphys1336} {\bibfield  {journal} {\bibinfo
  {journal} {Nat. Phys.}\ }\textbf {\bibinfo {volume} {5}},\ \bibinfo {pages}
  {555} (\bibinfo {year} {2009})}\BibitemShut {NoStop}%
\bibitem [{\citenamefont {Harriger}\ \emph {et~al.}(2011)\citenamefont
  {Harriger}, \citenamefont {Luo}, \citenamefont {Liu}, \citenamefont {Frost},
  \citenamefont {Hu}, \citenamefont {Norman},\ and\ \citenamefont
  {Dai}}]{P.Dai_DSHO}%
  \BibitemOpen
  \bibfield  {author} {\bibinfo {author} {\bibfnamefont {L.~W.}\ \bibnamefont
  {Harriger}}, \bibinfo {author} {\bibfnamefont {H.~Q.}\ \bibnamefont {Luo}},
  \bibinfo {author} {\bibfnamefont {M.~S.}\ \bibnamefont {Liu}}, \bibinfo
  {author} {\bibfnamefont {C.}~\bibnamefont {Frost}}, \bibinfo {author}
  {\bibfnamefont {J.~P.}\ \bibnamefont {Hu}}, \bibinfo {author} {\bibfnamefont
  {M.~R.}\ \bibnamefont {Norman}}, \ and\ \bibinfo {author} {\bibfnamefont
  {P.}~\bibnamefont {Dai}},\ }\href {\doibase 10.1103/PhysRevB.84.054544}
  {\bibfield  {journal} {\bibinfo  {journal} {Phys. Rev. B}\ }\textbf {\bibinfo
  {volume} {84}},\ \bibinfo {pages} {054544} (\bibinfo {year}
  {2011})}\BibitemShut {NoStop}%
\bibitem [{\citenamefont {Baron}(2009)}]{IXS_DSHO}%
  \BibitemOpen
  \bibfield  {author} {\bibinfo {author} {\bibfnamefont {A.}~\bibnamefont
  {Baron}},\ }\href@noop {} {\bibfield  {journal} {\bibinfo  {journal} {Journal
  of The Spectroscopical Society of Japan}\ }\textbf {\bibinfo {volume} {58}},\
  \bibinfo {pages} {205} (\bibinfo {year} {2009})}\BibitemShut {NoStop}%
\bibitem [{\citenamefont {Chernyshev}\ and\ \citenamefont
  {Zhitomirsky}(2006)}]{Chernychev_prl}%
  \BibitemOpen
  \bibfield  {author} {\bibinfo {author} {\bibfnamefont {A.~L.}\ \bibnamefont
  {Chernyshev}}\ and\ \bibinfo {author} {\bibfnamefont {M.~E.}\ \bibnamefont
  {Zhitomirsky}},\ }\href {\doibase 10.1103/PhysRevLett.97.207202} {\bibfield
  {journal} {\bibinfo  {journal} {Phys. Rev. Lett.}\ }\textbf {\bibinfo
  {volume} {97}},\ \bibinfo {pages} {207202} (\bibinfo {year}
  {2006})}\BibitemShut {NoStop}%
\bibitem [{\citenamefont {Chernyshev}\ and\ \citenamefont
  {Zhitomirsky}(2009)}]{Chernychev_prb}%
  \BibitemOpen
  \bibfield  {author} {\bibinfo {author} {\bibfnamefont {A.~L.}\ \bibnamefont
  {Chernyshev}}\ and\ \bibinfo {author} {\bibfnamefont {M.~E.}\ \bibnamefont
  {Zhitomirsky}},\ }\href {\doibase 10.1103/PhysRevB.79.144416} {\bibfield
  {journal} {\bibinfo  {journal} {Phys. Rev. B}\ }\textbf {\bibinfo {volume}
  {79}},\ \bibinfo {pages} {144416} (\bibinfo {year} {2009})}\BibitemShut
  {NoStop}%
\bibitem [{\citenamefont {Diallo}\ \emph {et~al.}(2009)\citenamefont {Diallo},
  \citenamefont {Antropov}, \citenamefont {Perring}, \citenamefont {Broholm},
  \citenamefont {Pulikkotil}, \citenamefont {Ni}, \citenamefont {Bud’ko},
  \citenamefont {Canfield}, \citenamefont {Kreyssig}, \citenamefont {Goldman},\
  and\ \citenamefont {McQueeney}}]{CaFe2As2_gamma}%
  \BibitemOpen
  \bibfield  {author} {\bibinfo {author} {\bibfnamefont {S.~O.}\ \bibnamefont
  {Diallo}}, \bibinfo {author} {\bibfnamefont {V.~P.}\ \bibnamefont
  {Antropov}}, \bibinfo {author} {\bibfnamefont {T.~G.}\ \bibnamefont
  {Perring}}, \bibinfo {author} {\bibfnamefont {C.}~\bibnamefont {Broholm}},
  \bibinfo {author} {\bibfnamefont {J.~J.}\ \bibnamefont {Pulikkotil}},
  \bibinfo {author} {\bibfnamefont {N.}~\bibnamefont {Ni}}, \bibinfo {author}
  {\bibfnamefont {S.~L.}\ \bibnamefont {Bud’ko}}, \bibinfo {author}
  {\bibfnamefont {P.~C.}\ \bibnamefont {Canfield}}, \bibinfo {author}
  {\bibfnamefont {A.}~\bibnamefont {Kreyssig}}, \bibinfo {author}
  {\bibfnamefont {A.~I.}\ \bibnamefont {Goldman}}, \ and\ \bibinfo {author}
  {\bibfnamefont {R.~J.}\ \bibnamefont {McQueeney}},\ }\href {\doibase
  10.1103/PhysRevLett.102.187206} {\bibfield  {journal} {\bibinfo  {journal}
  {Phys. Rev. Lett.}\ }\textbf {\bibinfo {volume} {102}},\ \bibinfo {pages}
  {187206} (\bibinfo {year} {2009})}\BibitemShut {NoStop}%
\bibitem [{\citenamefont {Ibuka}\ \emph {et~al.}(2017)\citenamefont {Ibuka},
  \citenamefont {Itoh}, \citenamefont {Yokoo},\ and\ \citenamefont
  {Endoh}}]{FeMn_gamma}%
  \BibitemOpen
  \bibfield  {author} {\bibinfo {author} {\bibfnamefont {S.}~\bibnamefont
  {Ibuka}}, \bibinfo {author} {\bibfnamefont {S.}~\bibnamefont {Itoh}},
  \bibinfo {author} {\bibfnamefont {T.}~\bibnamefont {Yokoo}}, \ and\ \bibinfo
  {author} {\bibfnamefont {Y.}~\bibnamefont {Endoh}},\ }\href {\doibase
  10.1103/PhysRevB.95.224406} {\bibfield  {journal} {\bibinfo  {journal} {Phys.
  Rev. B}\ }\textbf {\bibinfo {volume} {95}},\ \bibinfo {pages} {224406}
  (\bibinfo {year} {2017})}\BibitemShut {NoStop}%
\bibitem [{\citenamefont {Adams}\ \emph {et~al.}(2000)\citenamefont {Adams},
  \citenamefont {Mason}, \citenamefont {Fawcett}, \citenamefont {Menshikov},
  \citenamefont {Frost}, \citenamefont {Forsyth}, \citenamefont {Perring},\
  and\ \citenamefont {Holden}}]{FeGe2_gamma}%
  \BibitemOpen
  \bibfield  {author} {\bibinfo {author} {\bibfnamefont {C.~P.}\ \bibnamefont
  {Adams}}, \bibinfo {author} {\bibfnamefont {T.~E.}\ \bibnamefont {Mason}},
  \bibinfo {author} {\bibfnamefont {E.}~\bibnamefont {Fawcett}}, \bibinfo
  {author} {\bibfnamefont {A.~Z.}\ \bibnamefont {Menshikov}}, \bibinfo {author}
  {\bibfnamefont {C.~D.}\ \bibnamefont {Frost}}, \bibinfo {author}
  {\bibfnamefont {J.~B.}\ \bibnamefont {Forsyth}}, \bibinfo {author}
  {\bibfnamefont {T.~G.}\ \bibnamefont {Perring}}, \ and\ \bibinfo {author}
  {\bibfnamefont {T.~M.}\ \bibnamefont {Holden}},\ }\href {\doibase
  10.1088/0953-8984/12/39/311} {\bibfield  {journal} {\bibinfo  {journal} {J.
  Phys.: Condens. Matter}\ }\textbf {\bibinfo {volume} {12}},\ \bibinfo {pages}
  {8487} (\bibinfo {year} {2000})}\BibitemShut {NoStop}%
\bibitem [{\citenamefont {Luo}\ \emph {et~al.}(2020)\citenamefont {Luo},
  \citenamefont {Marcus}, \citenamefont {Trump}, \citenamefont {Kindervater},
  \citenamefont {Stone}, \citenamefont {Rodriguez-Rivera}, \citenamefont {Qiu},
  \citenamefont {McQueen}, \citenamefont {Tchernyshyov},\ and\ \citenamefont
  {Broholm}}]{twomag_DOS}%
  \BibitemOpen
  \bibfield  {author} {\bibinfo {author} {\bibfnamefont {Y.}~\bibnamefont
  {Luo}}, \bibinfo {author} {\bibfnamefont {G.~G.}\ \bibnamefont {Marcus}},
  \bibinfo {author} {\bibfnamefont {B.~A.}\ \bibnamefont {Trump}}, \bibinfo
  {author} {\bibfnamefont {J.}~\bibnamefont {Kindervater}}, \bibinfo {author}
  {\bibfnamefont {M.~B.}\ \bibnamefont {Stone}}, \bibinfo {author}
  {\bibfnamefont {J.~A.}\ \bibnamefont {Rodriguez-Rivera}}, \bibinfo {author}
  {\bibfnamefont {Y.}~\bibnamefont {Qiu}}, \bibinfo {author} {\bibfnamefont
  {T.~M.}\ \bibnamefont {McQueen}}, \bibinfo {author} {\bibfnamefont
  {O.}~\bibnamefont {Tchernyshyov}}, \ and\ \bibinfo {author} {\bibfnamefont
  {C.}~\bibnamefont {Broholm}},\ }\href@noop {} {\bibfield  {journal} {\bibinfo
   {journal} {arXiv:2002.06283}\ } (\bibinfo {year} {2020})}\BibitemShut
  {NoStop}%
\bibitem [{\citenamefont {Qin}\ \emph {et~al.}(2015)\citenamefont {Qin},
  \citenamefont {Zakeri}, \citenamefont {Ernst}, \citenamefont {Sandratskii},
  \citenamefont {Buczek}, \citenamefont {Marmodoro}, \citenamefont {Chuang},
  \citenamefont {Zhang},\ and\ \citenamefont {Kirschner}}]{Stoner_DOS}%
  \BibitemOpen
  \bibfield  {author} {\bibinfo {author} {\bibfnamefont {H.~J.}\ \bibnamefont
  {Qin}}, \bibinfo {author} {\bibfnamefont {K.}~\bibnamefont {Zakeri}},
  \bibinfo {author} {\bibfnamefont {A.}~\bibnamefont {Ernst}}, \bibinfo
  {author} {\bibfnamefont {L.~M.}\ \bibnamefont {Sandratskii}}, \bibinfo
  {author} {\bibfnamefont {P.}~\bibnamefont {Buczek}}, \bibinfo {author}
  {\bibfnamefont {A.}~\bibnamefont {Marmodoro}}, \bibinfo {author}
  {\bibfnamefont {T.~H.}\ \bibnamefont {Chuang}}, \bibinfo {author}
  {\bibfnamefont {Y.}~\bibnamefont {Zhang}}, \ and\ \bibinfo {author}
  {\bibfnamefont {J.}~\bibnamefont {Kirschner}},\ }\href {\doibase
  10.1038/ncomms7126} {\bibfield  {journal} {\bibinfo  {journal} {Nat.
  Commun.}\ }\textbf {\bibinfo {volume} {6}},\ \bibinfo {pages} {6126}
  (\bibinfo {year} {2015})}\BibitemShut {NoStop}%
\bibitem [{\citenamefont {Leong}\ \emph {et~al.}(2014)\citenamefont {Leong},
  \citenamefont {Lee}, \citenamefont {Lv},\ and\ \citenamefont
  {Phillips}}]{Stoner_intensity}%
  \BibitemOpen
  \bibfield  {author} {\bibinfo {author} {\bibfnamefont {Z.}~\bibnamefont
  {Leong}}, \bibinfo {author} {\bibfnamefont {W.-C.}\ \bibnamefont {Lee}},
  \bibinfo {author} {\bibfnamefont {W.}~\bibnamefont {Lv}}, \ and\ \bibinfo
  {author} {\bibfnamefont {P.}~\bibnamefont {Phillips}},\ }\href {\doibase
  10.1103/PhysRevB.90.125158} {\bibfield  {journal} {\bibinfo  {journal} {Phys.
  Rev. B}\ }\textbf {\bibinfo {volume} {90}},\ \bibinfo {pages} {125158}
  (\bibinfo {year} {2014})}\BibitemShut {NoStop}%
\bibitem [{\citenamefont {Mourigal}\ \emph {et~al.}(2013)\citenamefont
  {Mourigal}, \citenamefont {Fuhrman}, \citenamefont {Chernyshev},\ and\
  \citenamefont {Zhitomirsky}}]{DynamicalSF_Martin}%
  \BibitemOpen
  \bibfield  {author} {\bibinfo {author} {\bibfnamefont {M.}~\bibnamefont
  {Mourigal}}, \bibinfo {author} {\bibfnamefont {W.~T.}\ \bibnamefont
  {Fuhrman}}, \bibinfo {author} {\bibfnamefont {A.~L.}\ \bibnamefont
  {Chernyshev}}, \ and\ \bibinfo {author} {\bibfnamefont {M.~E.}\ \bibnamefont
  {Zhitomirsky}},\ }\href {\doibase 10.1103/PhysRevB.88.094407} {\bibfield
  {journal} {\bibinfo  {journal} {Phys. Rev. B}\ }\textbf {\bibinfo {volume}
  {88}},\ \bibinfo {pages} {094407} (\bibinfo {year} {2013})}\BibitemShut
  {NoStop}%
\bibitem [{\citenamefont {Sukhanov}\ \emph {et~al.}(2019)\citenamefont
  {Sukhanov}, \citenamefont {Pavlovskii}, \citenamefont {Bourges},
  \citenamefont {Walker}, \citenamefont {Manna}, \citenamefont {Felser},\ and\
  \citenamefont {Inosov}}]{Mn3Ge_PRB}%
  \BibitemOpen
  \bibfield  {author} {\bibinfo {author} {\bibfnamefont {A.~S.}\ \bibnamefont
  {Sukhanov}}, \bibinfo {author} {\bibfnamefont {M.~S.}\ \bibnamefont
  {Pavlovskii}}, \bibinfo {author} {\bibfnamefont {P.}~\bibnamefont {Bourges}},
  \bibinfo {author} {\bibfnamefont {H.~C.}\ \bibnamefont {Walker}}, \bibinfo
  {author} {\bibfnamefont {K.}~\bibnamefont {Manna}}, \bibinfo {author}
  {\bibfnamefont {C.}~\bibnamefont {Felser}}, \ and\ \bibinfo {author}
  {\bibfnamefont {D.~S.}\ \bibnamefont {Inosov}},\ }\href {\doibase
  10.1103/PhysRevB.99.214445} {\bibfield  {journal} {\bibinfo  {journal} {Phys.
  Rev. B}\ }\textbf {\bibinfo {volume} {99}},\ \bibinfo {pages} {214445}
  (\bibinfo {year} {2019})}\BibitemShut {NoStop}%
\bibitem [{\citenamefont {Chen}\ \emph {et~al.}(2020)\citenamefont {Chen},
  \citenamefont {Gaudet}, \citenamefont {Dasgupta}, \citenamefont {Marcus},
  \citenamefont {Lin}, \citenamefont {Chen}, \citenamefont {Tomita},
  \citenamefont {Ikhlas}, \citenamefont {Zhao}, \citenamefont {Chen},
  \citenamefont {Stone}, \citenamefont {Tchernyshyov}, \citenamefont
  {Nakatsuji},\ and\ \citenamefont {Broholm}}]{Mn3Ge_arxiv}%
  \BibitemOpen
  \bibfield  {author} {\bibinfo {author} {\bibfnamefont {Y.}~\bibnamefont
  {Chen}}, \bibinfo {author} {\bibfnamefont {J.}~\bibnamefont {Gaudet}},
  \bibinfo {author} {\bibfnamefont {S.}~\bibnamefont {Dasgupta}}, \bibinfo
  {author} {\bibfnamefont {G.~G.}\ \bibnamefont {Marcus}}, \bibinfo {author}
  {\bibfnamefont {J.}~\bibnamefont {Lin}}, \bibinfo {author} {\bibfnamefont
  {T.}~\bibnamefont {Chen}}, \bibinfo {author} {\bibfnamefont {T.}~\bibnamefont
  {Tomita}}, \bibinfo {author} {\bibfnamefont {M.}~\bibnamefont {Ikhlas}},
  \bibinfo {author} {\bibfnamefont {Y.}~\bibnamefont {Zhao}}, \bibinfo {author}
  {\bibfnamefont {W.~C.}\ \bibnamefont {Chen}}, \bibinfo {author}
  {\bibfnamefont {M.~B.}\ \bibnamefont {Stone}}, \bibinfo {author}
  {\bibfnamefont {O.}~\bibnamefont {Tchernyshyov}}, \bibinfo {author}
  {\bibfnamefont {S.}~\bibnamefont {Nakatsuji}}, \ and\ \bibinfo {author}
  {\bibfnamefont {C.}~\bibnamefont {Broholm}},\ }\href@noop {} {\bibfield
  {journal} {\bibinfo  {journal} {arXiv:2001.09495}\ } (\bibinfo {year}
  {2020})}\BibitemShut {NoStop}%
\bibitem [{\citenamefont {Kampf}(1996)}]{Itinerant_Hubbard}%
  \BibitemOpen
  \bibfield  {author} {\bibinfo {author} {\bibfnamefont {A.~P.}\ \bibnamefont
  {Kampf}},\ }\href {\doibase 10.1103/PhysRevB.53.747} {\bibfield  {journal}
  {\bibinfo  {journal} {Phys. Rev. B}\ }\textbf {\bibinfo {volume} {53}},\
  \bibinfo {pages} {747} (\bibinfo {year} {1996})}\BibitemShut {NoStop}%
\bibitem [{\citenamefont {Blöchl}(1994)}]{S1}%
  \BibitemOpen
  \bibfield  {author} {\bibinfo {author} {\bibfnamefont {P.~E.}\ \bibnamefont
  {Blöchl}},\ }\href {\doibase 10.1103/PhysRevB.50.17953} {\bibfield
  {journal} {\bibinfo  {journal} {Phys. Rev. B}\ }\textbf {\bibinfo {volume}
  {50}},\ \bibinfo {pages} {17953} (\bibinfo {year} {1994})}\BibitemShut
  {NoStop}%
\bibitem [{\citenamefont {Kresse}\ and\ \citenamefont {Joubert}(1999)}]{S2}%
  \BibitemOpen
  \bibfield  {author} {\bibinfo {author} {\bibfnamefont {G.}~\bibnamefont
  {Kresse}}\ and\ \bibinfo {author} {\bibfnamefont {D.}~\bibnamefont
  {Joubert}},\ }\href {\doibase 10.1103/PhysRevB.59.1758} {\bibfield  {journal}
  {\bibinfo  {journal} {Phys. Rev. B}\ }\textbf {\bibinfo {volume} {59}},\
  \bibinfo {pages} {1758} (\bibinfo {year} {1999})}\BibitemShut {NoStop}%
\bibitem [{\citenamefont {Perdew}\ \emph {et~al.}(1996)\citenamefont {Perdew},
  \citenamefont {Burke},\ and\ \citenamefont {Ernzerhof}}]{S3}%
  \BibitemOpen
  \bibfield  {author} {\bibinfo {author} {\bibfnamefont {J.~P.}\ \bibnamefont
  {Perdew}}, \bibinfo {author} {\bibfnamefont {K.}~\bibnamefont {Burke}}, \
  and\ \bibinfo {author} {\bibfnamefont {M.}~\bibnamefont {Ernzerhof}},\ }\href
  {\doibase 10.1103/PhysRevLett.77.3865} {\bibfield  {journal} {\bibinfo
  {journal} {Phys. Rev. Lett.}\ }\textbf {\bibinfo {volume} {77}},\ \bibinfo
  {pages} {3865} (\bibinfo {year} {1996})}\BibitemShut {NoStop}%
\bibitem [{\citenamefont {Kresse}\ and\ \citenamefont {Hafner}(1993)}]{S4}%
  \BibitemOpen
  \bibfield  {author} {\bibinfo {author} {\bibfnamefont {G.}~\bibnamefont
  {Kresse}}\ and\ \bibinfo {author} {\bibfnamefont {J.}~\bibnamefont
  {Hafner}},\ }\href {\doibase 10.1103/PhysRevB.47.558} {\bibfield  {journal}
  {\bibinfo  {journal} {Phys. Rev. B}\ }\textbf {\bibinfo {volume} {47}},\
  \bibinfo {pages} {558} (\bibinfo {year} {1993})}\BibitemShut {NoStop}%
\bibitem [{\citenamefont {Kresse}\ and\ \citenamefont {Hafner}(1994)}]{S5}%
  \BibitemOpen
  \bibfield  {author} {\bibinfo {author} {\bibfnamefont {G.}~\bibnamefont
  {Kresse}}\ and\ \bibinfo {author} {\bibfnamefont {J.}~\bibnamefont
  {Hafner}},\ }\href {\doibase 10.1103/PhysRevB.49.14251} {\bibfield  {journal}
  {\bibinfo  {journal} {Phys. Rev. B}\ }\textbf {\bibinfo {volume} {49}},\
  \bibinfo {pages} {14251} (\bibinfo {year} {1994})}\BibitemShut {NoStop}%
\bibitem [{\citenamefont {Kresse}\ and\ \citenamefont
  {Furthmüller}(1996{\natexlab{a}})}]{S6}%
  \BibitemOpen
  \bibfield  {author} {\bibinfo {author} {\bibfnamefont {G.}~\bibnamefont
  {Kresse}}\ and\ \bibinfo {author} {\bibfnamefont {J.}~\bibnamefont
  {Furthmüller}},\ }\href {\doibase
  https://doi.org/10.1016/0927-0256(96)00008-0} {\bibfield  {journal} {\bibinfo
   {journal} {Computational Materials Science}\ }\textbf {\bibinfo {volume}
  {6}},\ \bibinfo {pages} {15} (\bibinfo {year}
  {1996}{\natexlab{a}})}\BibitemShut {NoStop}%
\bibitem [{\citenamefont {Kresse}\ and\ \citenamefont
  {Furthmüller}(1996{\natexlab{b}})}]{S7}%
  \BibitemOpen
  \bibfield  {author} {\bibinfo {author} {\bibfnamefont {G.}~\bibnamefont
  {Kresse}}\ and\ \bibinfo {author} {\bibfnamefont {J.}~\bibnamefont
  {Furthmüller}},\ }\href {\doibase 10.1103/PhysRevB.54.11169} {\bibfield
  {journal} {\bibinfo  {journal} {Phys. Rev. B}\ }\textbf {\bibinfo {volume}
  {54}},\ \bibinfo {pages} {11169} (\bibinfo {year}
  {1996}{\natexlab{b}})}\BibitemShut {NoStop}%
\bibitem [{\citenamefont {Grechnev}\ \emph {et~al.}(2009)\citenamefont
  {Grechnev}, \citenamefont {Fedorchenko}, \citenamefont {Logosha},
  \citenamefont {Panfilov}, \citenamefont {Svechkarev}, \citenamefont
  {Filippov}, \citenamefont {Lyashchenko},\ and\ \citenamefont
  {Evdokimova}}]{S8}%
  \BibitemOpen
  \bibfield  {author} {\bibinfo {author} {\bibfnamefont {G.~E.}\ \bibnamefont
  {Grechnev}}, \bibinfo {author} {\bibfnamefont {A.~V.}\ \bibnamefont
  {Fedorchenko}}, \bibinfo {author} {\bibfnamefont {A.~V.}\ \bibnamefont
  {Logosha}}, \bibinfo {author} {\bibfnamefont {A.~S.}\ \bibnamefont
  {Panfilov}}, \bibinfo {author} {\bibfnamefont {I.~V.}\ \bibnamefont
  {Svechkarev}}, \bibinfo {author} {\bibfnamefont {V.~B.}\ \bibnamefont
  {Filippov}}, \bibinfo {author} {\bibfnamefont {A.~B.}\ \bibnamefont
  {Lyashchenko}}, \ and\ \bibinfo {author} {\bibfnamefont {A.~V.}\ \bibnamefont
  {Evdokimova}},\ }\href {\doibase
  https://doi.org/10.1016/j.jallcom.2009.03.123} {\bibfield  {journal}
  {\bibinfo  {journal} {Journal of Alloys and Compounds}\ }\textbf {\bibinfo
  {volume} {481}},\ \bibinfo {pages} {75} (\bibinfo {year} {2009})}\BibitemShut
  {NoStop}%
\bibitem [{\citenamefont {Togo}\ and\ \citenamefont {Tanaka}(2015)}]{S9}%
  \BibitemOpen
  \bibfield  {author} {\bibinfo {author} {\bibfnamefont {A.}~\bibnamefont
  {Togo}}\ and\ \bibinfo {author} {\bibfnamefont {I.}~\bibnamefont {Tanaka}},\
  }\href {\doibase https://doi.org/10.1016/j.scriptamat.2015.07.021} {\bibfield
   {journal} {\bibinfo  {journal} {Scripta Materialia}\ }\textbf {\bibinfo
  {volume} {108}},\ \bibinfo {pages} {1} (\bibinfo {year} {2015})}\BibitemShut
  {NoStop}%
\bibitem [{\citenamefont {Dewhurst}\ \emph {et~al.}()\citenamefont {Dewhurst},
  \citenamefont {Sharma}, \citenamefont {Nordström}, \citenamefont {Cricchio},
  \citenamefont {Granäs},\ and\ \citenamefont {Gross}}]{S10}%
  \BibitemOpen
  \bibfield  {author} {\bibinfo {author} {\bibfnamefont {J.~K.}\ \bibnamefont
  {Dewhurst}}, \bibinfo {author} {\bibfnamefont {S.}~\bibnamefont {Sharma}},
  \bibinfo {author} {\bibfnamefont {L.}~\bibnamefont {Nordström}}, \bibinfo
  {author} {\bibfnamefont {F.}~\bibnamefont {Cricchio}}, \bibinfo {author}
  {\bibfnamefont {O.}~\bibnamefont {Granäs}}, \ and\ \bibinfo {author}
  {\bibfnamefont {E.~K.~U.}\ \bibnamefont {Gross}},\ }\href
  {http://elk.sourceforge.net} {\enquote {\bibinfo {title} {All-electron
  full-potential linearised augmented-plane-wave (fp-lapw) package,
  http://elk.sourceforge.net},}\ }\BibitemShut {NoStop}%
\bibitem [{\citenamefont {Perdew}\ \emph {et~al.}(1992)\citenamefont {Perdew},
  \citenamefont {Chevary}, \citenamefont {Vosko}, \citenamefont {Jackson},
  \citenamefont {Pederson}, \citenamefont {Singh},\ and\ \citenamefont
  {Fiolhais}}]{S11}%
  \BibitemOpen
  \bibfield  {author} {\bibinfo {author} {\bibfnamefont {J.~P.}\ \bibnamefont
  {Perdew}}, \bibinfo {author} {\bibfnamefont {J.~A.}\ \bibnamefont {Chevary}},
  \bibinfo {author} {\bibfnamefont {S.~H.}\ \bibnamefont {Vosko}}, \bibinfo
  {author} {\bibfnamefont {K.~A.}\ \bibnamefont {Jackson}}, \bibinfo {author}
  {\bibfnamefont {M.~R.}\ \bibnamefont {Pederson}}, \bibinfo {author}
  {\bibfnamefont {D.~J.}\ \bibnamefont {Singh}}, \ and\ \bibinfo {author}
  {\bibfnamefont {C.}~\bibnamefont {Fiolhais}},\ }\href@noop {} {\bibfield
  {journal} {\bibinfo  {journal} {Physical review B}\ }\textbf {\bibinfo
  {volume} {46}},\ \bibinfo {pages} {6671} (\bibinfo {year}
  {1992})}\BibitemShut {NoStop}%
\end{thebibliography}%

\end{document}